\documentclass[nohyper,12pt,letterpaper]{JHEP3}
\usepackage[dvips]{epsfig}
\usepackage{amsfonts,amssymb}

\def\crampest{\medmuskip = 1mu plus 1mu minus 1mu}

\def\ben{\begin{equation}}
\def\een{\end{equation}}

\let\a=\alpha    

\let\C=\Chi

\def\nn{\nonumber} \def\bd{\begin{document}} \def\ed{\end{document}}
\def\ds{\documentstyle} \let\fr=\frac \let\bl=\bigl \let\br=\bigr
\let\Br=\Bigr \let\Bl=\Bigl
\let\bm=\bibitem
\let\na=\nabla
\let\pa=\partial \let\ov=\overline
\newcommand{\be}{\begin{equation}}
\newcommand{\ee}{\end{equation}}
\def\ba{\begin{array}}
\def\ea{\end{array}}
\def\ft#1#2{{\textstyle{{\scriptstyle #1}\over {\scriptstyle #2}}}}
\def\fft#1#2{{#1 \over #2}}
\def\del{\partial}
\def\vp{\varphi}
\def\sst#1{{\scriptscriptstyle #1}}
\def\oneone{\rlap 1\mkern4mu{\rm l}}
\def\td{\tilde}
\def\wtd{\widetilde}
\def\ie{{\it i.e.\ }}
\def\eg{{\it e.g.\ }}
\def\dalemb#1#2{{\vbox{\hrule height .#2pt
        \hbox{\vrule width.#2pt height#1pt \kern#1pt
                \vrule width.#2pt}
        \hrule height.#2pt}}}
\def\square{\mathord{\dalemb{6.8}{7}\hbox{\hskip1pt}}}
\newcommand{\ho}[1]{$\, ^{#1}$}
\newcommand{\hoch}[1]{$\, ^{#1}$}
\newcommand{\bea}{\begin{eqnarray}}
\newcommand{\eea}{\end{eqnarray}}
\newcommand{\ra}{\rightarrow}
\newcommand{\lra}{\longrightarrow}
\newcommand{\Lra}{\Leftrightarrow}
\newcommand{\tr}{{\rm tr} }
\newcommand{\Tr}{{\rm Tr} }
\def\0{{\sst{(0)}}}
\def\1{{\sst{(1)}}}
\def\2{{\sst{(2)}}}
\def\3{{\sst{(3)}}}
\def\4{{\sst{(4)}}}
\def\5{{\sst{(5)}}}
\def\6{{\sst{(6)}}}
\def\7{{\sst{(7)}}}
\def\8{{\sst{(8)}}}
\def\n{{\sst{(n)}}}
\def\cA{{{\cal A}}}
\def\cF{{{\cal F}}}
\def\tV{\widetilde V}
\def\tW{\widetilde W}
\def\tH{\widetilde H}
\def\tE{\widetilde E}
\def\tF{\widetilde F}
\def\tA{\widetilde A}
\def\im{{{\rm i}}}
\def\tY{{{\wtd Y}}}
\def\ep{{\epsilon}}
\def\vep{{\varepsilon}}
\def\R{\rlap{\rm I}\mkern3mu{\rm R}}
\def\bD{{{\bar D}}}

\def\R{\rlap{\rm I}\mkern3mu{\rm R}}
\def\bD{{{\bar D}}}
\def\R{{{\mathbb R}}}
\def\C{{{\mathbb C}}}
\def\H{{{\mathbb H}}}
                   
\def\CP{{{\mathbb C}{\mathbb P}}}
\def\RP{{{\mathbb R}{\mathbb P}}}
\def\Z{{{\mathbb Z}}}
\def\bA{{{\mathbb A}}}
\def\bB{{{\mathbb B}}}
\def\bC{{{\mathbb C}}}
\def\bR{{{\mathbb R}}}
\def\bD{{{\mathbb D}}}
\def\bE{{{\mathbb E}}}
\def\bZ{{{\mathbb Z}}}
\def\bI{{{\mathbb I}}}
\def\Re{{{\frak{Re}}}}
\def\Im{{{\frak{Im}}}}
\def\cosec{{\,\hbox{cosec}\,}}
\def\Gm{{\Gamma_{\!\! -}}}
\def\Gp{{\Gamma_{\!\! +}}}
\def\stan{{standard }}
\def\nonstan{{supernumerary }}

\def\cosech{{\hbox{cosech}}}

\def\etcyc{{\hbox{and cyclic}}}
\def\btheta{{\bar\theta}}
\def\alp{{\alpha'}^3}

\newcommand{\tamphys}{\it Center for Theoretical Physics,
Texas A\&M University, College Station, TX 77843, USA}

\newcommand{\mitchell}{\it George P. \& Cynthia W.
Mitchell Institute for Fundamental Physics,\\
Texas A\&M University, College Station, TX 77843-4242, USA}
\newcommand{\umich}{\it Michigan Center for Theoretical Physics,
University of Michigan\\ Ann Arbor, MI 48109, USA}
\newcommand{\upenn}{\it Department of Physics and Astronomy,
University of Pennsylvania, Philadelphia,  PA 19104, USA}
\newcommand{\SISSA}{\it  SISSA-ISAS and INFN, Sezione di Trieste\\
Via Beirut 2-4, I-34013, Trieste, Italy}

\newcommand{\newton}{\it Isaac Newton Institute for Mathematical
Sciences,\\
20 Clarkson Road,  University of Cambridge,
Cambridge CB3 0EH, UK}

\newcommand{\ihp}{\it Institut Henri Poincar\'e\\
  11 rue Pierre et Marie Curie, F 75231 Paris Cedex 05}

\newcommand{\damtp}{\it DAMTP, Centre for Mathematical Sciences,
 Cambridge University\\  Wilberforce Road, Cambridge CB3 OWA, UK}
\newcommand{\itp}{\it Institute for Theoretical Physics, University of
California\\ Santa Barbara, CA 93106, USA}

\newcommand{\imperial}{\it The Blackett Laboratory,
Imperial College London\\
Prince Consort Road, London SW7 2AZ. }

\newcommand{\icrea}{\it Instituci\'o Catalana de Recerca i
Estudis Avan\c cats,\\
Departament ECM, Facultat de F\'\i sica, \\
Universitat de Barcelona, Diagonal 647,\\
E-08028 Barcelona, Spain.}

\newcommand{\auth}{
H. L\"u\hoch{\ddagger1},
C.N. Pope\hoch{\ddagger1}, K.S. Stelle\hoch{\star2} and
P.K. Townsend\hoch{\clubsuit3}}

\title{\Large\bf String and M-theory  Deformations of Manifolds with 
Special Holonomy}

\author{ H. L\"u and C.N. Pope\thanks{Research supported in part by 
DOE grant
DE-FG03-95ER40917}\\
\mitchell}

\author{K.S. Stelle\thanks{Research supported in part by the EC under 
TMR contracts HPRN-CT-2000-00131 and MRTN-CT-2004-005104 and by PPARC
under rolling grant PPA/G/O/2002/00474}\\
\imperial}

\author{P.K. Townsend\\
                   \damtp}

\abstract{ The $R^4$-type corrections to ten and eleven dimensional
supergravity required by string and M-theory imply corrections to
supersymmetric supergravity compactifications on manifolds of special
holonomy, which deform the metric away from the original
holonomy. Nevertheless, in many such cases, including Calabi-Yau
compactifications of string theory and $G_2$-compactifications of
M-theory, it has been shown that the deformation preserves
supersymmetry because of associated corrections to the supersymmetry
transformation rules, Here, we consider Spin(7) compactifications in
string theory and M-theory, and a class of non-compact $SU(5)$
backgrounds in M-theory. Supersymmetry survives in all these cases
too, despite the fact that the original special holonomy is perturbed
into general holonomy in each case.}

\keywords{Strings, Supersymmetry, Holonomy, M-theory}

\preprint{MIFP-04-19\\ Imperial/TP/041002\\ DAMTP-2004-114\\ 
\tt{hep-th/0410176}}

\begin{document}

\section{Introduction}
                   
An important question in superstring theory is whether there are
compactifications to a lower-dimensional Minkowski spacetime that
preserve some fraction of the supersymmetry of the 10-dimensional
Minkowski vacuum. At string tree-level, this question can be addressed
within the $\alpha'$ expansion of the effective supergravity theory;
we shall consider only type II string theories for which the leading
$\alpha'$ correction occurs at order $\alp$ and has an $R^4$
structure.  If fermions are omitted, then the corrected action for the
metric and dilaton takes the form 
\be
\label{lageff} {\cal L} =
\sqrt{-g}\, e^{-2\phi}\, \left( R + 4(\del\phi)^2 - c\, \alp\,
Y\right)\label{genlag} 
\ee
for a known constant $c$, proportional to $\zeta(3)$, and  a known  scalar $Y$
that is quartic in the Riemann tensor of the 10-dimensional spacetime. 

We shall primarily be concerned with solutions to the equations of motion of this
action for which the dilaton is constant to lowest order and the
10-dimensional spacetime is the product of 2-dimensional Minkowski
spacetime with some initially Ricci-flat Riemannian 8-dimensional manifold
$M_8$, with curvature tensor $R_{ijk\ell}$. In this case,
\bea
Y  &=& \ft1{64} (t^{i_1\cdots i_8}\, t^{j_1\cdots j_8} - \ft14
\ep^{i_1\cdots i_8}\, \ep^{j_1\cdots j_8})\, 
R_{i_1 i_2 j_1 j_2}\, R_{i_3 i_4 j_3 j_4}\,
   R_{i_5 i_6 j_5 j_6}\, R_{i_7 i_8 j_7 j_8}\,.\label{Y8}\\
&\equiv& Y_0 - Y_2\,,\label{y0y2}
\eea
where the $SO(8)$-invariant $t$-tensor is defined by 
\be
t^{i_1\cdots i_8} \, M_{i_1 i_2}\dots M_{i_7 i_8} =
   24 M_{i}{}^{j}\,  M_{j}{}^{k}\,M_{k}{}^{\ell}\,
M_{\ell}{}^{i} - 6 (M_{i}{}^{j}\, M_{j}{}^{i})^2
\ee
for an arbitrary antisymmetric tensor $M_{i_1 i_2}$.  Note that
in the decomposition $Y=Y_0-Y_2$ in (\ref{y0y2}), the subscripts 
on $Y_0$ and $Y_2$ indicate that these are the terms built with
0 and 2 epsilon tensors respectively.

If $M_8$ is assumed to be compact, then consistency with the 
initially nondilatonic structure
requires
\cite{grosswitten, g2mod} 
\be
\int_{M_8} Y= {\cal O}\left(\alpha'\right).  
\ee 
The simplest way to
satisfy this criterion is to demand that $Y=0$ to leading order in the
$\alpha'$ expansion, and this is satisfied if $M_8=K_8$ for some
manifold $K_8$ of special holonomy (which is necessarily
Ricci-flat). This is also what one needs for the lowest-order solution
to preserve supersymmetry. The number of supersymmetries preserved
equals the number of linearly-independent Killing spinors; i.e., real
$SO(8)$ spinors $\psi_0$ satisfying 
\be
\label{killingspinor}
R_{ijk\ell}\tilde\Gamma^{k\ell}\psi_0 =0.  
\ee 
where $\tilde\Gamma^i$
are the $16\times 16$ real $SO(8)$ Dirac matrices (the notation is
chosen to agree with that of \cite{g2mod}). 

    To see why one has $Y=0$ when $M_8=K_8$ of special holonomy, we 
note that $Y$ can be
expressed as a Berezin integral \cite{grosswitten} 
\be 
Y \propto \int d^{16}\psi\,
\exp\left[\left(\bar\psi_- \tilde\Gamma^{ij}\,
\psi_-\right)\left( \bar\psi_+\tilde\Gamma^{k\ell}\, \psi_+\right)\,
R_{ijk\ell}\right] \, , \label{YBer} 
\ee
where $\bar\psi=\psi^T$ and the integration is over the 16 components
of a real anticommuting constant $SO(8)$ spinor or, equivalently, over
all 16 linearly-independent $SO(8)$ spinors $\psi$.  We can write
$\psi=\psi_+ +\psi_-$, where $\psi_\pm$ are the chiral and antichiral
projections of $\psi$.  If
there are any Killing spinors amongst them, as there will be if
$M_8=K_8$ of special holonomy, then the rules of Berezin integration 
imply that $Y=0$.

  If we use $\a$ and $\dot\a$ to denote 8-component right-handed and
left-handed spinor indices respectively, then up to an inessential
constant factor, (\ref{YBer}) can be rewritten as
\bea
Y &=& \ep^{\a_1\cdots \a_8}\, \ep^{\dot\beta_1\cdots \dot\beta_8}\,
    \Gamma^{i_1 i_2}_{\a_1\a_2} \cdots \Gamma^{i_7 i_8}_{\a_7\a_8}\,
    \Gamma^{j_1 j_2}_{\dot\beta_1\dot\beta_2}\cdots
\Gamma^{j_7 j_8}_{\dot\beta_7\dot\beta_8}\times \nn\\
&& R_{i_1 i_2 j_1 j_2}\, R_{i_3 i_4 j_3 j_4}\,
   R_{i_5 i_6 j_5 j_6}\, R_{i_7 i_8 j_7 j_8}\,.
\eea
It is straightforward to show that
\bea
\ft1{256} \ep^{\a_1\cdots \a_8}\, \Gamma^{i_1 i_2}_{\a_1\a_2}\,
\Gamma^{i_3 i_3}_{\a_3\a_3}\,
    \Gamma^{i_6 i_6}_{\a_5\a_6}\, \Gamma^{i_7 i_8}_{\a_7\a_8}&\equiv&
t_+^{i_1\cdots i_8} =
t^{i_1\cdots i_8} + \ft12 \ep^{i_1\cdots i_8}\,,\label{tplus}\\
\ft1{256} \ep^{\dot\beta_1\cdots \dot\beta_8}\,
\Gamma^{j_1 j_2}_{\dot\beta_1\dot\beta_2}\,
    \Gamma^{j_3 j_4}_{\dot\beta_3\dot\beta_4}\,
    \Gamma^{j_5 j_6}_{\dot\beta_5\dot\beta_6}\,
    \Gamma^{j_7 j_8}_{\dot\beta_7\dot\beta_8} &\equiv&
t_-^{i_1\cdots i_8} = 
t^{j_1\cdots j_8} - \ft12 \ep^{j_1\cdots j_8}\,,\label{tminus}
\eea
(with one overall convention choice determining which right-hand side has the
plus sign, and which the minus).  Thus we see that (\ref{YBer})
is of the form $t_-\, t_+\, R^4$, and hence gives rise
to (\ref{Y8}).

   It might appear from this result that configurations with constant
dilaton and a spacetime of the form $\bE^{(1,1)}\times K_8$ will
automatically continue to be solutions of the $\alp$-corrected field
equations, in which case one would expect the special holonomy of
$K_8$ to guarantee that supersymmetry is preserved. This is true
if $K_8 = T^4\times K_4$ for a 4-manifold of $SU(2)$ holonomy (i.e., a
hyper-K\"ahler manifold) but it is false in general because, although $Y$
vanishes for a manifold of special holonomy, its {\it variation} with
respect to the metric yields a tensor $X_{ij}$ as a source in the
corrected Einstein equations, and this tensor may be non-zero even
though $Y=0$.  Specifically, under the circumstances described, the
corrected Einstein and dilaton equations are to ${\cal O}(\alp)$
\bea
\label{einstmod}
R_{ij} +2  \nabla_i\nabla_j\, \phi   &=& c\, \alp\, X_{ij}\,, \nn\\
R + 4 \nabla^2\phi  &=& 0\,.
\eea

    When $K_8= T^2\times K_6$ for a 6-dimensional manifold of $SU(3)$
holonomy; i.e., when $K_6$ is a Calabi-Yau (CY) manifold, the correction
due to the tensor $X_{ij}$ deforms the leading-order CY metric to one
of $U(3)$ holonomy \cite{freemanpope}.  However, as shown in
\cite{cfpss}, this deformation does not break the supersymmetry of the
undeformed solution because there is a compensating $\alp$ correction
to the gravitino supersymmetry transformation law or, equivalently, to
the covariant derivative acting on spinors. More precisely, it was
shown that there is a {\it possible} corrected covariant derivative
that has this property; it is expected that this will be needed for a
construction of the supersymmetric extension of the Lagrangian
(\ref{lageff}), but this complete construction has yet to be carried out in
sufficient detail. Nevertheless, this perspective makes it clear that
any proposed corrections to the supersymmetry transformations must be
expressible in purely Riemannian terms, {\it without} the use of any
special structures arising from special holonomy. The proposal of
\cite{cfpss} passes this test, which is quite non-trivial in view of
the fact that the methods used (details of which can be found in
the review \cite{fpssmeudon}) relie heavily on the K\"ahler properties of CY
manifolds. It turns out that the purely Riemannian form of the
corrected covariant derivative has an obvious extension to
8-manifolds, which is all that will be needed here, and the result can
then be summarised by saying that the standard covariant derivative
$\nabla_i$ acting on $SO(8)$ spinors must be replaced by
\be\label{Rcovderiv}
\hat\nabla_i = \nabla_i - {3c\over4} \alp \, 
\left[ (\nabla^j\, R_{ik m_1 m_2})\, R_{j\ell m_3 m_4}\, 
R^{k\ell}{}_{m_5 m_6}\right]\, \tilde\Gamma^{m_1 \cdots m_6} + 
{\cal O}\left(\alpha'{}^4\right)
\ee
It is important to appreciate that it was {\it not} claimed in
\cite{cfpss} that this is the only correction of relevance to this
order in the $\alpha'$ expansion, but rather that this term is
sufficient for lowest-order backgrounds of the form $\bE^{(1,3)}\times
K_6$ and its related toroidal compactifications such as
$\bE^{(1,1)}\times T^2\times K_6$. In particular, there could be
additional terms that are non-zero for a spacetime of the form
$\bE^{(1,1)}\times M_8$ but which vanish when $M_8=T^2\times K_6$.

   It is also important to appreciate that the question of whether or
not the special-holonomy backgrounds continue to be supersymmetric in
the face of $\alp$ corrections is one that cannot be
addressed unless one has knowledge of the order $\alp$ 
correction to the gravitino
transformation rule\footnote{After the completion of the first version
of this paper, but before its submission to the archives, 
there appeared a paper \cite{constantin} having some overlap with 
our Spin(7) results, but without any discussion of the  
order $\alp$ corrections to the supersymmetry transformation rules 
that are needed to address the question that is central here; i.e.,
whether supersymmetry is maintained in the corrected background.}.
At perturbation orders higher than
$\alp$, there will also certainly be further corrections. In the present paper,
however, we limit the discussion to at most this order.

    Similar issues arise when $K_8= S^1\times K_7$ for a 7-manifold
$K_7$ of $G_2$ holonomy, as one would expect since the special case
$K_7=S^1\times K_6$ yields $K_8= T^2\times K_6$.  In particular, the
$\alp R^4$ corrections to supergravity arising from the exchange of
massive string states must deform any lowest-order compactification on
a manifold of initial $G_2$ holonomy to a compactification on a manifold of
generic $SO(7)$ holonomy, and it is far from obvious that such a 
solution will continue to
preserve supersymmetry. Moreover, as $G_2$-manifolds are not K\"ahler, 
the methods used to
address this issue in the CY case are no longer available. However, using the
existence of the associative 3-form on a $G_2$ manifold, we were able
to show in a previous paper \cite{g2mod} that there is a simple
correction to the covariant derivative on spinors that implies
supersymmetry preservation of the modified solution, and we used this
result to determine the explicit form of the correction for most of
the known classes of cohomogeneity-one 7-metrics with $G_2$ structures
(as was done for an analogous class of CY metrics in \cite{lupost}).
Despite the fact that our simple form of the corrected covariant
derivative made explicit use of the associative 3-form available only
for $G_2$ manifolds, it was again found possible (by making crucial
use of properties of $G_2$ manifolds) to rewrite this corrected
covariant derivative in purely Riemannian terms. There is again an
obvious extension to 8-manifolds, and the resulting covariant
derivative acting on $SO(8)$ spinors was once again found to be
(\ref{Rcovderiv}).

    Corrections to the effective supergravity action of the form $R^4$
arise not only at tree level in string theory but also at the one-loop
level. This correction is related by dualities to an analogous $R^4$
M-theory correction to 11-dimensional supergravity. The latter has a
structure that differs from the $R^4$ tree-level string-theory
correction, and it also includes an $A\wedge X_8$ Chern-Simons (CS)
term that is absent at tree level in string theory. However, for
$G_2$ compactifications, these differences are unimportant, so we were
able to lift our string-theory results directly to M-theory. There was
a subtlety, however, arising from the fact that an $\alp$ correction to the
dilaton was needed at tree-level in string theory whereas there is no
dilaton in 11 dimensions. However, the effect of the dilaton in string
theory can be achieved in M-theory by a modification of the $R^4$
invariant via a field redefinition. We were thus able to show (i) that
M-theory implies a modification of $G_2$ compactifications of
11-dimensional supergravity in which the 7-metric of $G_2$ holonomy is
deformed to one of generic, $SO(7)$, holonomy, and (ii) that (${\cal
N} =1$) supersymmetry of the effective four-dimensional theory is
maintained, despite this deformation, at least to order $\alp$.

   One purpose of this paper is to extend our results on $G_2$
compactifications, as summarised above, to Spin(7)
compactifications. In this respect this should be considered as a
companion paper to \cite{g2mod}. At tree-level in string
theory our Spin(7) results are similar to those obtained in
\cite{g2mod}, although there are some additional technical
difficulties and subtleties.  We also determine explicit
supersymmetry-preserving $\alp$ corrections for some of the known
classes of cohomogeneity-one 8-metrics with Spin(7) structures. At
one-loop in string theory, or in M-theory, however, there are more substantial
differences arising from the necessity to take into account the
Chern-Simons terms associated with the $R^4$ corrections, and for compact
$K_8$ there is also a topological constraint that must be taken into
account. We find that there is nevertheless a supersymmetric 
deformation of Spin(7) compactifications of  M-theory, and hence
of 1-loop corrected IIA superstring theory, whether or not the Spin(7) 
manifold is actually compact. 

  Another purpose of this paper is to consider the effects of the
$R^4$ corrections of M-theory on compactifications of
eleven-dimensional supergravity on ten-manifolds of $SU(5)$ holonomy.
This is of considerable interest because it probes aspects of M-theory
lying beyond those that are accessible from perturbative string
theory.  We find corrections to the leading-order backgrounds, and we
also consider their supersymmetry. As for the Spin(7)
compactifications, there is a topological constraint to take into
account. This constraint arises for any $SU(5)$-holonomy 10-manifold
$K_{10}$ with non-trivial homology group $H_8$. As $H_8$ is isomorphic
to $H_2$ for any compact 10-manifold (by Poincar\'e duality) and since
$H_2$ is obviously non-trivial (because $K_{10}$ is K\"ahler), there
is a topological constraint on $SU(5)$-holonomy compactifications of
M-theory, and this constraint even applies to non-compact backgrounds
if $H_8$ is non-trivial. The implications for supersymmetry of of this
topological constraint are not at present fully clear to us, so we
shall restrict ourselves here to the class of non-compact 10-manifolds
$K_{10}$ of $SU(5)$ holonomy for which $H_8$ is trivial and for which
the topological constraint is therefore trivially satisfied.  Even so,
our results for this case are worthy of note; we find that the same
correction to the gravitino transformation rule that ensured the
continued supersymmetry of the Spin(7) holonomy backgrounds also
implies that the corrected $SU(5)$ holonomy backgrounds maintain
supersymmetry.  Interestingly, the corrected $SU(5)$ background is no
longer even K\"ahler, but it is still a complex manifold of vanishing
first Chern class.

\section{Spin(7) Preliminaries}\label{spin7sec}

    As pointed out in \cite{g2mod}, the structure of the $R^4$
invariant $Y$ implies that the tensor $X_{ij}$, which arises from the
variation of $Y$ and which appears in the corrected Einstein field equation,
takes the form
\be\label{Xdecom}
X_{ij} = \tilde X_{ij} + \nabla^k\nabla^\ell X_{ijk\ell}
\ee
for a tensor $\tilde X_{ij}$, quartic in the curvatures and a tensor
$X_{ijk\ell}$ that is cubic in curvatures.  We will show in this
section that if the variational expression $X_{ij}$ is then 
evaluated in a background that has Spin(7) holonomy, then  
\be
\tilde X_{ij} =0
\ee
and in fact $X_{ij}$ is given by 
\be
X_{ij} = \ft12 c^{mnk}{}_{(i}\, c^{pq\ell}{}_{j)} \, \nabla_k\, \nabla_\ell\, 
    Z_{mnpq} + \nabla^k\, \nabla_\ell\, Z_{mnk(i}\, c^{mn\ell}{}_{j)}\,,
\label{xijres0}
\ee
where $c_{ijk\ell}$ is the calibrating 4-form on the Spin(7) holonomy 
background, and 
\be
Z^{mnpq} =\ft1{64}\, \ep^{mn i_1\cdots i_6}\, \ep^{pq j_1\cdots j_6}\, 
R_{i_1 i_2 j_1 j_2}\, R_{i_3 i_4 j_3 j_4}\, R_{i_5 i_6 j_5 j_6}\,.
\label{zmnpq0}
\ee

\subsection{Properties of Spin(7) manifolds}
\label{subsec.spin7}

We begin with some basic results about Spin(7) manifolds. There is a
single real (commuting) Killing spinor, $\eta$, that is either chiral or
anti-chiral. We choose conventions in which $\eta$ is anti-chiral,
corresponding to the Spin(7) decomposition
\be
8_+ \longrightarrow 8\,,\qquad 8_-\longrightarrow 7+1\,.\label{8pm} 
\ee
of the chiral/anti-chiral spinor irreps of $SO(8)$. Note that the
vector representation of SO(8) remains irreducible:
\be
8_v\longrightarrow 8\, .\label{8v}
\ee

     We shall normalise the commuting Killing spinor $\eta$ so 
that $\bar\eta\, \eta=1$
(where $\bar\eta = \eta^T$). Given this normalisation, and introducing
$\Gamma_9$ as the (real) $SO(8)$ chirality matrix, we have the
identities
\be
\Gamma_i\, \eta\, \bar\eta\,\Gamma^i = \oneone_+\,, \qquad
\eta\, \bar\eta -\ft18 \Gamma_{ij}\, \eta\, \bar\eta\, \Gamma^{ij}
=\oneone_-
\ee
where
\be
\oneone_\pm  \equiv \ft12(\oneone \pm \Gamma_9), 
\ee
which is the identity operator projected into the chiral or
anti-chiral spin bundle.

   The calibrating 4-form  has components that are expressible as
\be\label{cal4}
c_{ijk\ell} = \bar\eta\, \Gamma_{ijk\ell}\, \eta\,.\label{c4def}
\ee
It is straightforward to establish the following identities:
\bea
c_{ijk\ell}\, \Gamma^{k\ell}\, \eta &=& -6 \Gamma_{ij}\, \eta\,,
\label{cids1}\\
c_{ijk p}\, c^{\ell mnp} &=& 6 \delta_{ijk}^{\ell mn} - 36
        \delta_{[i}{}^{[\ell}\, c_{jk]}{}^{mn]}\,.\label{cids3}
\eea
Recalling the Killing spinor integrability condition
$R_{ijk\ell}\tilde\Gamma^{k\ell}\eta=0$, one can also show that
\be
R_{ijk\ell}\, c^{k\ell}{}_{mn} = 2 R_{ijmn}\,;\label{cids2}
\ee  
this is the condition for Spin(7) holonomy.
 
\subsection{Correction to the Einstein equations}\label{achtsec}

   In order to derive the $\alp$ corrections to the Einstein equations
at string tree level, we need to evaluate the variation of the
quartic-curvature term $Y$.  This was relatively straightforward in
the case of corrections to six-dimensional Calabi-Yau backgrounds
$K_6$, \cite{freemanpope}, and for corrections to seven-dimensional
$G_2$-holonomy backgrounds $K_7$ \cite{g2mod}.  The reason for this is
that in each case, one has $SO(8)$ Killing spinors of both chiralities
in the $K_8=\R^2\times K_6$, or $K_8=\R\times K_7$ eight-dimensional
transverse space.  This means that when one varies the metrics or
vielbeins in the Berezin integral (\ref{YBer}), the only terms that
can survive are those where the metrics in one of the Riemann tensors
itself are varied.  This is because they are the only terms where one
does not inevitably end up with Killing spinors linked to an
unvaried Riemann tensor and thus vanishing by virtue of
(\ref{killingspinor}).  

   Additionally, because of the non-chiral nature of the
Killing spinors in $K_8$ in the previous cases, it was straightforward
to express the variation of $Y$, originally written in terms of spinors in
the Berezin integral (\ref{YBer}),  in terms of tensorial quantities built 
from Riemann tensors and the K\"ahler form of $K_6$ or the associative
3-form of $K_7$.  This stemmed from the fact that for both chiral and
antichiral $SO(8)$ spinors, one had decompositions under $SU(3)$ or 
$G_2$ that provided a one-to-one mapping between the vector and the
spinor representation in $K_6$ or $K_7$.  

   In the case of Spin(7) holonomy manifolds $K_8$ things are more
subtle for two reasons.  Firstly, we have a Killing spinor of only one
eight-dimensional chirality, which we are taking, by convention
choice, to be antichiral.  This means that we could, {\it a priori},
encounter non-vanishing terms in the variation of $Y$, defined in
(\ref{YBer}), in which vielbeins used in contracting the Riemann tensors
onto the Dirac matrices are varied, leaving all four Riemann tensors
unvaried.

   Secondly, we can see from (\ref{8pm}) and (\ref{8v}) that, while the
$8_+$ spinor representation of $SO(8)$ is indeed isomorphic to the
$8_v$ vector representation in a Spin(7) background, the $8_-$ spinor
representation is not.  This could lead to obstacles in rewriting the
variation of $Y$, given by (\ref{YBer}), in a purely tensorial form.

    To address these problems, it is helpful to introduce two further
quartic-curvature invariants, which we shall call $Y_-$ and $Y_+$.  These
are defined in terms of Berezin integrals analogous to (\ref{YBer}), 
except that now we have
\bea 
Y_+ &\propto& \int d^{8}\psi_+\, d^8\chi_+\,
\exp\left[\left(\bar\psi_+ \tilde\Gamma^{ij}\,
\psi_+\right)\left( \bar\chi_+\tilde\Gamma^{k\ell}\, \chi_+\right)\,
R_{ijk\ell}\right] \,. \label{YBerp}\\
Y_- &\propto& \int d^{8}\psi_-\, d^8\chi_-\,
\exp\left[\left(\bar\psi_- \tilde\Gamma^{ij}\,
\psi_-\right)\left( \bar\chi_-\tilde\Gamma^{k\ell}\, \chi_-\right)\,
R_{ijk\ell}\right] \,, \label{YBerm}
\eea
The integration in (\ref{YBerp}) is over two independent sets of 
chiral $SO(8)$  
spinors, while in (\ref{YBerm}) it is over independent two sets of 
antichiral spinors.  
 From (\ref{tplus}) and (\ref{tminus}), we see that $Y_+$ and $Y_-$ are
given by 
\bea
Y_\pm &=& \ft1{64} t_\pm^{i_1\cdots i_8}\, t_\pm^{j_1\cdots j_8}
R_{i_1 i_2 j_1 j_2}\cdots R_{i_7 i_8 j_7 j_8}\,,\nn\\
&=& \ft1{64} (t^{i_1\cdots i_8}\, t^{j_1\cdots j_8} 
    \pm t^{i_1\cdots i_8}\,  \ep^{j_1\cdots j_8}
+  \ft14
\ep^{i_1\cdots i_8}\, \ep^{j_1\cdots j_8})\, 
R_{i_1 i_2 j_1 j_2} \cdots R_{i_7 i_8 j_7 j_8}\,,\label{Ypm}\\
&\equiv & Y_0 \pm Y_1 + Y_2\,,\label{yy0yy2}
\eea
where $Y_0$ and $Y_2$ are the same as in (\ref{y0y2}), and $Y_1$ is the term
in (\ref{Ypm}) that is linear in the epsilon tensor.

   A crucial property of the invariants $Y_\pm$ is that they differ
from the actual effective action contribution $Y$ by terms that are purely
topological in $D=8$:
\be
Y_\pm - Y =\pm Y_1 + 2 Y_2 =
(\pm t^{i_1\cdots i_8} + \ft12 \ep^{i_1\cdots i_8})
\ep^{j_1\cdots j_8}\, 
   R_{i_1i_2j_1j_2}\cdots R_{i_7i_8j_7j_8} \,. 
\ee
As an 8-form, written in terms of the curvature 2-forms $\Theta_{ij}$, 
the difference is given by
\be
{*(Y_\pm -Y)}= 16 (\pm t^{i_1\cdots i_8} + \ft12 \ep^{i_1\cdots i_8})\,
\Theta_{i_1 i_2}\wedge \cdots\wedge \Theta_{i_7 i_8}\,,
\ee
which makes the topological nature manifest.  Because of this, the
integrals of $Y$, $Y_+$ and $Y_-$ all have the same 
variation,\footnote{To be precise, when we say that
$Y$, $Y_+$ and $Y_-$ all have the same variation, we mean that their
variations differ by total derivatives.  At string tree level, where
these quantities are multiplied by $e^{-2\phi}$, the total derivatives
will integrate by parts to give contributions involving derivatives of
the dilaton when comparing the corrected Einstein equations. However,
since the $Y$ term is accompanied by an explicit $\alp$ factor, and since the
dilaton is constant in the leading-order background these extra derivative
terms contribute at best at order ${\alpha'}^6$ in the corrected Einstein
equations, and thus they may be neglected at the $\alp$ order to which we
are working.  At string one-loop, or in M-theory, there is no dilaton
prefactor, and so the integration by parts simply gives zero.}
evaluated on an eight-dimensional curved background, and so we can use
either $Y_+$ or $Y_-$ in place of $Y$ for the purpose of computing
the variation $X_{ij}$ (even though $Y_+$ does not vanish in the
special-holonomy background).\footnote{We should note, because a
failure to do so has caused some confusion in the earlier literature,
that the computed result for the Berezin integral for $Y$
that is given in Ref.\ \cite{grosswitten} is actually the result 
obtained by computing
the Berezin integral for $Y_+$, but since $Y_+=Y$ for the CY 
compactifications considered
there, and since the variations are also the same, the 
distinction was unimportant there. In
our case, however, the distinction is important.}

   Each of the $Y_\pm$ has its own advantages and disadvantages, when used
in place of $Y$ to calculate the variation $X_{ij}$.  If we vary
$Y_-$, then it is manifest that no terms from the variation of the 
bare vielbeins contracting Riemann tensors $R^\mu{}_{\nu\rho\sigma}$ onto Dirac
matrices $\Gamma^{ij}$ will survive in the Berezin integration.
This is because we will always have a contribution either of the form
$R_{ijk\ell}\, \Gamma^{k\ell}\,\eta$ or $\Gamma^{ij}\eta\,
R_{ijk\ell}$ in every term where the explicit vielbeins are varied,
and these then vanish by virtue of the integrability condition for the
(antichiral) Killing spinor.  Thus only terms arising from the
variation of metrics contained within the connections from
which $R^\mu{}_{\nu\rho\sigma}$ is composed will survive.
This means that, after integration by parts, the variation of $Y_-$
will necessarily involve only terms constructed from two covariant
derivatives acting on (Riemann)$^3$ structures, and that there will be no terms
quartic in Riemann tensors without derivatives.  The
drawback to using $Y_-$, however, is that there is no isomorphism between the
decompositions of the $8_-$ and $8_v$ representations of $SO(8)$ 
under restriction to
Spin(7), and therefore we do not have a simple direct way of re-expressing
$\delta Y_-$ in purely bosonic tensorial terms.

  On the other hand, if we vary $Y_+$ then the isomorphism between the
irreducible $8_+$ and $8_v$ representations of $SO(8)$ under
restriction to Spin(7) does in this case provide us with a simple way to recast
$\delta Y_+$ in purely bosonic tensorial terms.  The drawback to using
$Y_+$, however, is that there are no spinor zero modes at all in the
Berezin integral (\ref{YBerp}), and so it is not immediately manifest
that the terms coming from the variation of the bare vielbeins that
contract Riemann tensors $R^\mu{}_{\nu\rho\sigma}$ onto Dirac matrices
$\Gamma^{ij}$ will not contribute.  Indeed, $Y_+$ itself does not even
vanish in the Spin(7) background.

  We can however make use of the complementary properties that are 
manifested in the different expressions $Y$, $Y_+$ and $Y_-$, and 
thereby ``have our cake and eat it too.''  In particular, we note that
the difference $Y_+-Y_-$ is also topological,
\be
Y_+ - Y_- =  2t^{i_1\cdots i_8}\, 
\ep^{j_1\cdots j_8}\, 
   R_{i_1i_2j_1j_2}\cdots R_{i_7i_8j_7j_8} \,,\label{ydif}
\ee
which means that after the varied expression is specialised to a Spin(7) 
background, we must have it that $\delta Y_-$ and $\delta Y_+$ give the same
contribution to the corrected Einstein equations, at order $\alp$.  In
particular, we can see that (\ref{ydif}) may be written in terms of
Riemann tensors $R^\mu{}_{\nu\rho\sigma}$ without the use of any
bare metrics or vielbeins.  We can now invoke the above observation
that the variation of $Y_-$ does not contain any terms coming from 
the variation of bare vielbeins to see that there will be no such terms in 
the variation of $Y_+$ either.  Then, we are in a position to exploit the
isomorphism between the decompositions of the $8_+$ and $8_v$ representations
of $SO(8)$ under restriction to Spin(7) to obtain a simple tensorial 
expression for $\delta Y_+$, and hence $\delta Y$. 

    It follows from (\ref{tplus}) and (\ref{Ypm}) that we shall have
\bea
Y_+ &\propto & \ep^{\a_1\cdots \a_8}\, \ep^{\beta_1\cdots\beta_8}\,
    \Gamma^{i_1 i_2}_{\a_1\a_2}\cdots \Gamma^{i_7 i_8}_{\a_7\a_8}\,
    \Gamma^{j_1 j_2}_{\beta_1\beta_2}\cdots
  \Gamma^{j_7 j_8}_{\beta_7\beta_8}\times \nn\\
&& R_{i_1 i_2 j_1 j_2}\, R_{i_3 i_4 j_3 j_4}\,
   R_{i_5 i_6 j_5 j_6}\, R_{i_7 i_8 j_7 j_8}\,.\label{4rpp}
\eea
Because the $8_+$ and $8_v$ representation become the same
irreducible representation of Spin(7), the expression (\ref{4rpp})
can be rewritten such that only vector indices are needed.
Specifically, the mapping between $8_+$ and $8_v$ is implemented by
\be
\nu^i_\a = \Gamma^i_{\a\dot\beta}\, \eta^{\dot\beta}\,.
\ee
This matrix has unit determinant, and so we can write
\be
\ep^{\a_1\cdots \a_8} = \nu_{i_1}^{\a_1}\cdots \nu_{i_8}^{\a_8}\, 
\ep^{i_1\cdots i_8}\,.
\ee

   Since we have argued that there will be no contributions coming
from varying the bare vielbeins in (\ref{4rpp}), after specialising
the varied expression to a Spin(7) background, we need only vary the
metrics in the connections from which the Riemann tensors themselves
are constructed.  Up to a constant factor, which is as yet inessential
to our discussion, we therefore have 
\bea 
\delta Y_+ &=&
4\ep^{\a_1\cdots \a_8}\, \ep^{\beta_1\cdots \beta_8}\, (\Gamma^{i_1
i_2})_{\a_1\a_2}\cdots (\Gamma^{i_7 i_8})_{\a_7\a_8}\, (\Gamma^{j_1
j_2})_{\beta_1\beta_2}\cdots (\Gamma^{j_7 j_8})_{\beta_7\beta_8}\,
\times \nn\\
 && R_{i_1 i_2 j_1 j_2}\cdots R_{i_5 i_6 j_5 j_6}\,
\delta R_{i_7 i_8 j_7 j_8}\,, \nn\\
&=&  8\ep^{\a_1\cdots \a_8}\, \ep^{\beta_1\cdots \beta_8}\,
(\Gamma^{i_1 i_2})_{\a_1\a_2}\cdots (\Gamma^{i_7 i_8})_{\a_7\a_8}\,
(\Gamma^{j_1 j_2})_{\beta_1\beta_2}\cdots
   (\Gamma^{j_7 j_8})_{\beta_7\beta_8}\, \times \nn\\
 && R_{i_1 i_2 j_1 j_2}\cdots R_{i_5 i_6 j_5 j_6}\, 
\nabla_{i_7}\nabla_{j_7}\, \delta g_{i_8 j_8}\,.\label{epformvar}
\eea
where $\delta R_{i_7 i_8 j_7 j_8}$ denotes the variation of the Riemann
tensor with respect to the metric.

   From the properties (\ref{cids1}) and (\ref{cids3}), one easily shows that
\be
\bar\eta \, \Gamma_i\, \Gamma^{k\ell}\, \Gamma_j\, \eta = 
   c_{ij}{}^{k\ell} + 2 \delta_{ij}^{k\ell}\,,
\ee
and hence, using (\ref{cids2}) repeatedly, we see that up to a further
inessential overall factor (and specialised to the Spin(7) background) we have
\be
\delta Y_+ = Z^{mnpq}\, (c_{mn}{}^{ij}  + 2\delta^{ij}_{mn})\, 
(c_{pq}{}^{k\ell} + 2 \delta^{k\ell}_{pq})\, \delta R_{ijk\ell}
\ee
where
\be
Z^{mnpq} =\ft1{64}\, \ep^{mn i_1\cdots i_6}\, \ep^{pq j_1\cdots j_6}\, 
R_{i_1 i_2 j_1 j_2}\, R_{i_3 i_4 j_3 j_4}\, R_{i_5 i_6 j_5 j_6}\,.
\label{zmnpq}
\ee
The following useful properties of $Z^{mnpq}$ can easily be established:
\bea
&&Z^{mnpq}= Z^{pqmn}=-Z^{nmpq} = -Z^{mnqp}\,,\nn\\
&& \nabla_m\, Z^{mnpq}=0\,,\qquad 
c_{mnp r}\, Z^{mnpq}= 0\,.\label{zidents}
\eea
We therefore conclude that the variation of $Y$ gives
\be
X_{ij} = \ft12 c^{mnk}{}_{(i}\, c^{pq\ell}{}_{j)} \, \nabla_k\, \nabla_\ell\, 
    Z_{mnpq} + \nabla^k\, \nabla_\ell\, Z_{mnk(i}\, c^{mn\ell}{}_{j)}\,.
\label{xijres}
\ee
Note that a simple calculation 
using (\ref{cids1}), (\ref{cids3}), (\ref{cids2}) and (\ref{zidents}) 
shows that
\be
g^{ij}\, X_{ij} = \square Z\,,\label{trxij}
\ee
and hence from (\ref{einstmod}) we learn that
\bea
R_{ij} &=& c\, \alp\, (X_{ij} + \nabla_i\nabla_j\, Z)\,,\label{ricmod}\\
\phi &=& -\ft12 c\, \alp\, Z\,.
\eea

\section{Correction to the Supersymmetry Transformation Rule}
\label{susycorrectsec}

      Since the effect of the $\alp$ corrections is to deform the 
original Spin(7) metric to one that is no longer Ricci flat, it 
follows that it will no longer have Spin(7) holonomy and so it will
no longer admit a covariantly constant spinor.  However, one knows that
at the same time as the $\alp$ corrections to the string effective 
action set in, there also will be corresponding corrections to the 
supersymmetry transformation rules at the $\alp$ order.  These were
discussed in the context of six-dimensional Calabi-Yau backgrounds
in Refs \cite{cfpss,fpssmeudon}, where it was indeed shown that the deformed 
metrics, which acquire an extra $U(1)$ factor to their original
undeformed $SU(3)$ holonomy, have the feature of still admitting spinors
that are constant with respect to a modified covariant derivative.  
This $O(\alp)$
modification can be understood as the necessary correction to the 
gravitino transformation rule at this order.  This issue was discussed
further for Calabi-Yau backgrounds in \cite{lupost}, and for 
seven-dimensional backgrounds with $G_2$ holonomy in 
\cite{g2mod}.\footnote{It should be emphasised that if these order $\alp$
corrections to the supersymmetry transformation rule are not included,
then one will not be able to demonstrate the preservation of supersymmetry
in the $\alp$-corrected backgrounds.}

   Here, we shall begin by introducing the following modified covariant
derivative,\footnote{Note that for our present purposes, where we
are simply concerned with establishing the circumstances under which 
a Killing spinor exists, we may view two formulations of a
gravitino transformation rule as equivalent if they agree when acting 
on the putative Killing spinor.  Here, as in much of the previous
literature, we shall commonly adopt this viewpoint.}
\be
D_i \equiv  \nabla_i + Q_i = \nabla_i + \ft14 c\, \alp\, 
   c_{ijk\ell}\, \nabla^j\, Z^{k\ell mn}\, \Gamma_{mn}\,.
\label{Ddef}
\ee 
where the $Z$-tensor is the one defined in (\ref{zmnpq}). After
some algebra, which involves making extensive use of properties given
in subsection \ref{subsec.spin7}, one finds that the integrability
condition $[D_i, D_j]\, \eta=0$ for the existence of a spinor
satisfying $D_i\, \eta=0$ precisely implies that (\ref{ricmod})
holds.  This, therefore, is our candidate expression for the
modification to the gravitino transformation rule in an originally
Spin(7) background; $\delta\psi_i = D_i\, \ep$.

    As it stands, (\ref{Ddef}) is written using the special tensor
$c_{ijk\ell}$ specific to a Spin(7) background.  One knows, of course,
that the modified supersymmetry transformation rules (and also the
modified equations of motion) should all be expressible in fully
covariant Riemannian terms, making no use of additional invariant
tensors that exist only in special backgrounds.  This question has
been addressed for Calabi-Yau and $G_2$ backgrounds in the previous
literature \cite{cfpss,g2mod}, and indeed the candidate expressions
for the modified supersymmetry transformation rules that were written
down in \cite{cfpss,fpssmeudon} were fully Riemannian expressions that
were shown to be compatible with special forms written in K\"ahler
language.  In \cite{g2mod}, it was shown that the Riemannian
expressions in \cite{cfpss,fpssmeudon} were also compatible with a
special form written using the calibrating 3-form in a $G_2$
background.

   Here, we shall show that the modified derivative $D_i$ defined in 
(\ref{Ddef}) can be re-expressed without the use of the special tensor 
$c_{ijk\ell}$ of a Spin(7) background, and that in fact (\ref{Ddef}) is
nothing but the Spin(7) specialisation of the Riemannian results
conjectured in Refs \cite{cfpss,fpssmeudon}.

    To do this, it is useful first to note that we have
\bea
c_{ijk\ell}\, \ep^{k\ell i_1\cdots i_6} &=& 
  \bar\eta\, \Gamma_{ijk}\, \Gamma_\ell\, \eta\,  \ep^{k\ell i_1\cdots i_6} = 
  \bar\eta\, \Gamma_{ijk}\, \Gamma^{k i_i\cdots i_6}\eta\nn\,,\\
 &=& -4 \delta_{ij}^{[ i_1 i_2}\, c^{i_3 i_4 i_5 i_6]}\,,
\eea
and hence
\be
Q_i = \ft1{64}c\, \alp\, \delta_{ij}^{[i_1 i_2}\, 
    c^{i_3 i_4 i_5 i_6]}\, \ep^{mn j_1\cdots j_6}\, 
\nabla^j\, (R_{i_1 i_2 j_1 j_2}\, R_{i_3 i_4 j_3 j_4}\, 
  R_{i_5 i_6 j_5 j_6})\, \Gamma_{mn}
\ee
Since all the permutations of the indices $\{i_1\cdots i_6\}$ involve at least
one of the Riemann tensors having a double contraction with $c_{ijk\ell}$, 
it follows that we can make use (\ref{cids2}) and thereby absorb all
occurrences of this special tensor.  After performing the necessary
combinatoric manipulations, and some further simplifications using
the Bianchi identity for the Riemann tensor, we arrive at the result
\be
Q_i = -\ft34 c\,\alp\,  (\nabla^j\, R_{ik m_1 m_2})\, R_{j\ell m_3 m_4}\,
R^{k\ell}{}_{m_5 m_6}\, \Gamma^{m_1 \cdots m_6}\,.\label{riem1}
\ee
In this form, $Q_i$ can be recognised as precisely the same
modification to the Killing spinor condition that was proposed in
\cite{cfpss}.  In that case, the proposal was based on a
consideration of deformations from $SU(3)$ holonomy for six-dimensional
Calabi-Yau backgrounds.  It was also shown in \cite{g2mod} that the
more stringent conditions arising for $G_2$ backgrounds
lead to exactly the same modification to the Killing spinor
condition.  Here, we have shown that the yet more stringent conditions
of a Spin(7) background again yield the same result, confirming the
validity of the Riemannian expression (\ref{riem1}) that was conjectured
in \cite{cfpss}.

    Of course since a six-dimensional space of $SU(3)$ holonomy (times
a line or circle) is just a special case of a $G_2$ manifold, and a
seven-dimensional space of $G_2$ holonomy (times a line or circle) is
a special case of a Spin(7) manifold, it follows that our derivation
here encompasses the previous $SU(3)$ and $G_2$ results in \cite{cfpss}
and \cite{g2mod}.

\section{$\alp$ Corrections for Eight-Dimensional K\"ahler Metrics}

    An eight-dimensional Ricci-flat K\"ahler metric is a Spin(7) metric,
since its SU(4) holonomy is contained within Spin(7).  Specifically,
the embedding can be seen by examining the decomposition of the three
eight-dimensional representations of the SO(8) tangent-space group
first to Spin(7) and then to SU(4):

\bigskip

\centerline{
\begin{tabular}{cccccc}
SO(8)  & & Spin(7)&  &SU(4)\\ \hline
$8_+$ &$\longrightarrow$ &  8 & $\longrightarrow$ &  $4 + \overline 4$  \\
$8_-$ & $\longrightarrow$ & $7+1$ & $\longrightarrow$ & $6+1+1$\\
$8_v$ & $\longrightarrow$ & 8 & $\longrightarrow$ & $4 +\overline 4$\\
\end{tabular}}
\bigskip

    The two singlets in the decomposition of the $8_-$ under SU(4)
indicate that there are two covariantly-constant left-handed
Majorana-Weyl spinors, say $\eta_1$ and $\eta_2$, in the
SU(4)-holonomy metric, which we may normalise to $\bar\eta_A\, \eta_B =
\delta_{AB}$.  From these, we may define complex left-handed
spinors $\eta_\pm$ and $\bar\eta_\pm$ as
\be
\eta_\pm \equiv \ft1{\sqrt2}\, (\eta_1 \pm \im\, \eta_2)\,,\qquad
\bar\eta_\pm \equiv \ft1{\sqrt2}\, (\bar\eta_1 \pm \im\, \bar\eta_2)\,.
\ee
We shall then have
\bea
&& J_{ij} = \im\, \bar\eta_+\, \Gamma_{ij}\, \eta_- = \bar\eta_1\,
\Gamma_{ij}\, \eta_2\,,\qquad 3 J_{[ij}\, J_{k\ell]} =
\bar\eta_+\, \Gamma_{ijk\ell}\, \eta_-\,,
\nn\\
&& \Omega_{ijk\ell} = \bar\eta_+ \, \Gamma_{ijk\ell} \, \eta_+\,,\qquad
\overline\Omega_{ijk\ell} = \bar\eta_- \,\Gamma_{ijk\ell} \, \eta_-\,,
\label{jom}
\eea
where $J_{ij}$ is the K\"ahler form, and $\Omega_{ijk\ell}$ is the
holomorphic 4-form, with its complex conjugate $\overline\Omega_{ijk\ell}$.

   We may take the calibrating 4-form $c_{ijk\ell}$ of the SU(4) metric,
viewed as a Spin(7) metric, to be given by $c_{ijk\ell} = \bar\eta_1\,
\Gamma_{ijk\ell}\, \eta_1$.  It then follows from (\ref{jom}) that we
shall have

\be
c_{ijk\ell} =\ft12(\Omega_{ijk\ell} + \overline\Omega_{ijk\ell})
                       + 3 J_{[ij}\, J_{k\ell]}\,.
\ee
In a K\"ahler metric, the only non-vanishing components of the Riemann tensor
are ``mixed'' on both the first index-pair and the second index-pair. In
other words if the $i$ index on $R_{ijk\ell}$ is holomorphic then
$j$ must be antiholomorphic, and {\it vice versa}, with a similar
property for $k$ and $\ell$.  From the definition (\ref{zmnpq}) of
$Z^{mnpq}$, it then follows that this tensor must similarly be mixed
on its $mn$ indices and in its $pq$ indices.  From this, it follows
that
\be
\Omega_{ijmn}\, Z^{mnpq} =0\,,
\ee
together with similar relations following from symmetries and
from conjugation.  A K\"ahler metric also has the property that
\be
J_k{}^m\, J_\ell{}^n\, R_{ijmn}= R_{ijk\ell}\,,\label{jjriem}
\ee
together with the analogous property on the first index-pair.  These
expressions can be written more elegantly using the ``hat'' notation
introduce in \cite{fpss}, where, for any vector $V_i$, one defines
\be
V_{\hat i} \equiv J_i{}^j \, V_j\,.\label{hatdef}
\ee
Thus (\ref{jjriem}) becomes $R_{\hat i\hat j k\ell}= R_{ijk\ell}$.
 From  (\ref{zmnpq}), it therefore follows that
\be
Z^{\hat i \hat jpq} = Z^{ijpq}\,,\qquad
Z^{mn \hat i \hat j} = Z^{mnij}\,.
\ee

   Using the above results, it is now straightforward to show that the
expression for $X_{ij}$ that we obtained for a Spin(7) background
in (\ref{xijres}) reduces to
\be
X_{ij} = \ft12 \nabla_{\hat i}\, \nabla_{\hat j}\,
   (J_{mn}\, J_{pq}\, Z^{mnpq})
\ee
in an eight-dimensional Ricci-flat K\"ahler background.  After a little
further manipulation, we find that the result (\ref{ricmod})
for the $\alp$ correction to the Ricci-flatness condition in a 
Spin(7) background
reduces for an eight-dimensional Ricci-flat K\"ahler background to
the corrected condition
\be
R_{ij} = c\, \alp\, (\nabla_{\hat i}\, \nabla_{\hat j} +
\nabla_i\, \nabla_j)\, Z\,,
\ee
where, as before, we have defined $Z\equiv Z^{mn}{}_{mn}$.
This is in agreement with the standard result that one obtains from the
calculation of the supersymmetric sigma-model beta-function at four
loops.

   In a similar manner, we can specialise the Spin(7) correction term $Q_i$
in the spinor covariant derivative $D_i=\nabla_i + Q_i$ to the case of
an eight-dimensional Ricci-flat K\"ahler metric.  Using the properties
discussed above, we find that $Q_i$ defined in (\ref{Ddef}) reduces to
\be
Q_i = \ft14 c\, \alp\,\nabla_{\hat i}\, (J_{k\ell}\, Z^{k\ell mn})\, 
\Gamma_{mn}
\,.
\ee
It was shown in \cite{lupost} that when acting on a covariantly-constant
spinor in a K\"ahler background one has
\be
(\Gamma_{ij} + \Gamma_{\hat i\hat j})\, \eta
  = 2 \im\, J_{ij}\, \eta\,,
\ee
and hence it follows that when acting on $\eta$, the modified covariant
derivative in the deformed background reduces to
\bea
D_i\, \eta &=& \nabla_i \, \eta + \ft{\im}{4}\,c\, \alp\, \nabla_{\hat i}\, (
J_{k\ell}\, J_{mn}\, Z^{k\ell mn})\, \eta\,,\nn\\
&=& \nabla_i \, \eta + \ft{\im}{2}\,c\, \alp\, (\nabla_{\hat i} \, Z)\, \eta\,.
\label{qform}
\eea
This last expression agrees with the one given in Refs \cite{cfpss,lupost}.

\section{Explicit Examples}

\subsection{$S^7$ principal orbits}

   Following \cite{trig2}, we introduce left-invariant 1-forms $L_{AB}$ for the
group manifold $SO(5)$.  These satisfy $L_{AB}=-L_{BA}$, and
\be
dL_{AB} = L_{AC}\wedge L_{CB}\,.
\ee
The 7-sphere is then given by the coset $SO(5)/SU(2)_L$, where we take
the obvious $SO(4)$ subgroup of $SO(5)$, and write it (locally) as
$SU(2)_L\times SU(2)_R$.

    If we take the indices $A$ and $B$ in $L_{AB}$ to range over the
values $0\le A\le 4$, and split them as $A=(a,4)$, with $0\le a\le 3$,
then the $SO(4)$ subgroup is given by $L_{ab}$.  This is decomposed as
$SU(2)_L\times SU(2)_R$, with the two sets of $SU(2)$ 1-forms given by
the self-dual and anti-self-dual combinations:
\be
R_i = \ft12(L_{0i} + \ft12\ep_{ijk}\, L_{jk})\,,\qquad
L_i = \ft12(L_{0i} - \ft12\ep_{ijk}\, L_{jk})\,,
\ee
where $1\le i\le 3$.  Thus the seven 1-forms in the $S^7$ coset will be
\be
P_a \equiv L_{a4}\,,\qquad R_1\,,\qquad R_2\,,\qquad R_3\,.
\ee

   The most general cohomogeneity-one metric ansatz for these $S^7$
principal orbits is
\be
ds_8^2 = dt^2 + a_i^2\, R_i^2 + b^2\, P_a^2\,.\label{8metans1}
\ee
Several complete nonsingular Spin(7) metrics are contained within this
class, including the original asymptotically conical (AC) example found
in Refs \cite{brysal,gibpagpop}, which is uniaxial, $a_1=a_2=a_3$, and the 
family of asymptotically locally
conical (ALC) examples found in \cite{cglpspin7}, which are biaxial, with
(say) $a_1=a_2$.  

    In the natural orthonormal basis for (\ref{8metans1}), namely
\be
e^0=dt\, \qquad e^i= a_i\, R_i\,,\qquad e^a= b\, P_a\,,
\ee
the calibrating 4-form has components $c_{ijk\ell}$ given by
\bea
1&=&-c_{0123}=c_{0145}=c_{0167}=c_{0246}=-c_{0257}=c_{0347}=
c_{0356}\,,\nn\\
&=& c_{1247}=c_{1256}=-c_{1346}=c_{1357}=c_{2345}=c_{2367}=-c_{4567}\,,
\eea
where we have assigned explicit index values $i=1,2,3$ and
$a=4,5,6,7$.  It is now a straightforward mechanical exercise, 
most easily implemented by computer,  to solve
first for the covariantly-constant spinor $\eta$ in the unmodified
Spin(7) background, yielding first-order equations for the metric
functions $a_i$ and $b$, and then to find the $\alp$-corrected
first-order equations that follow from imposing $D_i\, \eta=0$, where
$D_i$ is given in (\ref{Ddef}).\footnote{Note that when we do this, we
assume that $\eta$ retains the identical form that it had in the
uncorrected Ricci-flat background.  The test of the validity of this
assumption is that the corrected first-order equations we obtain under
this assumption do indeed imply that the corrected second-order
Einstein equations (\ref{einstmod}) are satisfied.}  The first-order
equations in the general triaxial case are rather complicated, and are
not easily presentable in this paper.  Here, we shall just give our
results in the uniaxial special case, where the three metric functions
$a_i$ are set equal, $a_i=a$.  We then find that $a$ and $b$ must
satisfy
\be
\fft{\dot a}{a} = \fft1{a} - \fft{a}{2b^2} - c\, \alp\, \dot S_1\,,
\qquad
\fft{\dot b}{b} = \fft{3 a}{4 b^2} - c\, \alp\, \dot S_2\,,
\label{s7mod}
\ee
where $c$ is the usual constant that we introduced in (\ref{genlag}),
and
\bea
S_1 &=& \fft{64239 a^6  - 227052 a^4  b^2  + 269712 a^2  b^4
- 101440 b^6}{1064b^{12}}\,,\nn\\
S_2 &=& \fft{3(-4389 a^6  + 16821 a^4  b^2  - 20997 a^2  b^4
+ 8756 b^6 )}{133 b^{12}}\,.
\eea
We can integrate the equations (\ref{s7mod}) to give
\bea
b(r)^2 &=& \ft32 e^{-2c\,\alpha'^3\, \bar S_2(r)}
\int^{r} e^{2c\,\alpha'^3\, \bar S_2(r')}\, dr'\,,\nn\\
a(r)^2 &=& 2 b(r)^{-\ft43}\, e^{-c\,\alpha'^3\, (2\bar S_1(r) +
\ft43 \bar S_2(r))}\,\int^r b(r')^{\ft43}\,
e^{c\,\alpha'^3\, (2\bar S_1(r') +
\ft43 \bar S_2(r'))} dr'\,,
\eea
where the variable $r$ is defined by $dr = a\, dt$ and the bars on
$S_1$ and $S_2$ denote that these quantities are evaluated in the
leading-order background.

\subsection{$SU(3)/U(1)$ principal orbits}

     The cosets $SU(3)/U(1)$, known as Aloff-Wallach spaces $N(k,\ell)$, are
characterised by two integers $k$ and $\ell$, which define the embedding of
the $U(1)$ subgroup $h$ of $SU(3)$ matrices according to
\be
h = \hbox{diag}(e^{\im\, k\, \theta}, e^{\im\, \ell\, \theta}, e^{-\im\,
(k+\ell)\, \theta})\,.
\ee
If one defines $m=-k-\ell$, it is evident that there is an $S_3$ 
symmetry given by the permutations of 
$(k,\ell,-k-\ell)$.

   We define left-invariant 1-forms $L_A{}^B$ for $SU(3)$, where
$A=1,2,3$, $L_A{}^A=0$, $(L_A{}^B)^\dagger =L_B{}^A$ and $dL_A{}^B =
\im\, L_A{}^C\wedge L_C{}^B$, and introduce the combinations
\bea
&&\sigma \equiv L_1{}^3\,,\qquad \Sigma\equiv L_2{}^3\,,\qquad \nu\equiv
L_1{}^2\,,\nn\\
&&
\lambda\equiv  \sqrt2\, \cos\td\delta \, L_1{}^1 + \sqrt2\,
\sin\td\delta \, L_2{}^2\,,\label{ldefs}\\
&&Q\equiv -\sqrt2\, \sin\td\delta \, L_1{}^1 + \sqrt2\,
\cos\td\delta \, L_2{}^2\,,\nn
\eea
where $Q$ is taken to be the $U(1)$ generator lying outside the
$SU(3)/U(1)$ coset, and
\be
\fft{k}{\ell} = -\tan\td\delta\,.
\ee
Thus $\td\delta$ is restricted to an infinite discrete set of values.

     We shall follow \cite{cglpspin7} and  use real left-invariant 1-forms 
defined
by $\sigma=\sigma_1+\im\, \sigma_2$, $\Sigma=\Sigma_1 + \im\,
\Sigma_2$ and $\nu=\nu_1 +\im\, \nu_2$.  The cohomogeneity one metrics 
can then be written as
\be
ds_8^2 = dt^2 + a^2\, \sigma_i^2 + b^2\, \Sigma_i^2 + c^2\, \nu_i^2 +
f^2\, \lambda^2\,,\label{d8ansatz}
\ee
where $a$, $b$, $c$ and $f$ are functions of the radial coordinate
$t$.  Using the Killing spinor equations that we have derived in this paper,
we obtain the first-order equations for this system up to $\alpha'^3$ order,
given by
\bea
\fft{\dot a}{a} &=& \fft{b^2 +c^2 - a^2}{abc} - 
\fft{\sqrt2\, f\, \cos\td\delta}{a^2} - \alpha'^3\, K_1\,,\nn\\
\fft{\dot b}{b} &=& \fft{a^2 + c^2 - b^2}{abc} +
\fft{\sqrt2\,f\,\cos\td\delta}{b^2} - \alpha'^3\, K_2\,,\nn\\
\fft{\dot c}{c} &=& \fft{a^2 + b^2 - c^2}{abc} +
\fft{\sqrt2\, f\, (\cos\td\delta - \sin\td\delta)}{c^2} -
\alpha'^3\, K_3\,,\nn\\
\fft{\dot f}{f} &=& -\fft{\sqrt2\, f\, (\cos\td\delta -\sin\td\delta)}{c^2}
+\fft{\sqrt2\, f\, \cos\td\delta}{a^2} -
\fft{\sqrt2\,f\,\sin\td\delta}{b^2} - \alpha^3\, K_4\,,
\eea
where the $K_i$'s are polynomial functions in $a$, $b$, $c$ and $f$.
(We have temporarily absorbed the constant $c$ into $\alp$ in the
discussion of this example, to avoid confusion with the metric function
$c$.)
We have explicitly verified that these first-order equations satisfy
the generalised higher-order second-order Einstein equations.  Owing
to the complexity of the expressions for the $K_i$'s, we shall not
present their general form, but give only a certain specific example.

   Local solutions of the first-order equations for Spin(7) holonomy
exist for all values of $k$ and $\ell$ \cite{cglpspin7}.  In general these
have conical singularities, but in the special case $N(1,0)$, 
or its permutation-related cousins $N(0,1)$ or $N(1,-1)$, then the solution,
first found in \cite{sparks}, is complete and non-singular.  The solution
is given by
\be
\bar a=\sqrt{(r-1)(r+5)}\,,\quad \bar b=(r+1)\,,\quad
\bar c=\sqrt{r^2-9}\,,\quad \bar f=-\sqrt{\fft{9(r-3)(r+5)}{2(r+3)(r-1)}}\,,
\ee
where the coordinate $r$ is related to $t$ by $dt=h\,dr\equiv
-\ft{3}{\sqrt2}f^{-1}\, dr$.
Note that we use barred notation to denote the background variables.
For this specific metric, we find that
\crampest{
\bea
K_1&=&\fft{162(4r^8 - 13 r^7 - 83 r^6 -409 r^5 +
81 r^4 - 1351 r^3 - 3993 r^2 - 39955 r -97641}{h\,(r-1)^8 (r+3)^7}
\,,\nn\\
K_2&=&\fft{648(r+1)(r^6+6r^5 -18 r^4 -112
r^3 -91 r^2 +58 r - 5604)}{h\,(r-1)^7 (r+3)^7}\,,\nn\\
K_3&=&\fft{162(4r^8 + 77r^7 + 547 r^6 + 2297 r^5 +
7311r^4 + 19527 r^3 + 34761 r^2 + 69491 r -11135)}{h\,(r-1)^7 (r+3)^8}
\,,\nn\\
K_4&=& \fft{2592(r+1)(r^2+2r-43)(3r^4+12r^3 -
170r^2-364 r -1049)}{h\,(r-1)^8(r+3)^8}
\,.
\eea
}

\section{Deformation of Spin(7) Compactifications of M-theory}

   In this section, we now consider analogous corrections to an
initial (Minkowski)$_3\times K_8$ background in M-theory, which is
related by dimensional reduction to type IIA string theory at the one
string-loop level.  To begin, we give a general discussion of the
known correction terms in the M-theory effective action.

\subsection{Corrections to  (Minkowski)$_3\times K_8$ backgrounds}

The corrections to the $D=11$ bosonic Lagrangian, which correspond to
the lift of 1-loop corrections in the type IIA string,  take
the form
\be
{\cal L}_1= -\fft{\beta}{1152} (\hat Y + 2 \hat Y_2 +\cdots)\, 
{\hat *\oneone} + 
    \beta\, (2\pi)^4\, \hat A_\3\wedge \hat X_\8\,,\label{mmod}
\ee
where $\hat X_\8 $ is given by
\be
\hat X_\8 = \fft1{192\, (2\pi)^4}\, [ 
 \tr\, \hat \Theta^4 - \ft14 (\tr\, \hat \Theta^2)^2]\,,
\ee
and $\hat Y$ and $\hat Y_2$ are eleven-dimensional 
analogues of the ten-dimensional quantities $Y$ and $Y_2$ described in
section \ref{spin7sec}, but now with the summation index ranges extended
to 11 rather than 10 values.
In particular, $\hat Y_2$ is proportional to the
covariant generalisation of the eight-dimensional Euler integrand,
\be
\hat Y_2 = \ft{315}{2} \hat R^{[M_1M_2}{}_{M_1 M_2}\, \cdots 
    \hat R^{M_7 M_8]}{}_{M_7 M_8}\,.\label{y2exp}
\ee
The constant $\beta$ now takes on the r\^ole played by $\alp$ in string
theory, and we shall work to order $\beta$ in the subsequent
discussion.  

    The ellipses in (\ref{mmod}) represent terms that vanish by use of
the leading-order field equations, and which therefore can be adjusted
by choice of field variables.  These changes of variable do not, of
course, affect the physics, but they can be used to advantage in order
to make the discussion more elegant.  By adding a specific term of
this type, we shall be able to ensure that the corrected equations 
of motion describing the modification to the Spin(7) holonomy internal
space are the same as those that we found at tree-level in string theory.
To achieve this, we shall take the bracketed volume term in (\ref{mmod}) 
to be
\be
\hat W  = \hat Y + 2 \hat Y_2 - \hat R\, \hat Z\,,\label{wdef}
\ee
and so
\be
{\cal L}_1 = -\fft{\beta}{1152}\, \hat W\, {\hat*\oneone} 
+ (2\pi)^4\, \beta\, \hat A_\3\wedge 
\hat X_\8\,,\label{mmod2}
\ee
The additional $\hat R\, \hat Z$ term is introduced for convenience by a field
rededinition of the metric, as in \cite{g2mod}, to compensate for the 
absence of a dilaton in M-theory.  It does not change the physics, but it
renders the equations more elegant. 

    The variation $\delta \int \sqrt{-\hat g}\,
\hat Y\, d^{11} x \equiv \int \sqrt{-\hat g}\, \hat Y_{MN}\,
\delta \hat g^{MN}\, d^{11} x$ yields
\be
\hat Y_{\mu\nu}= 0\,,\qquad \hat Y_{ij} = X_{ij}\,,
\ee  
in the 3-dimensional spacetime and the internal 8-dimensional manifold
respectively, after imposing the leading-order (Minkowski)$_3\times M_8$
background conditions, where $M_8$ is a Spin(7) manifold.
The tensor $X_{ij}$ is given by (\ref{xijres}).  
Varying $\hat Y'\equiv (\hat Y - \hat R\, \hat Z)$ instead of $\hat Y$, 
we find 
\be
\hat Y'_{\mu\nu}= - g_{\mu\nu}\, \square Z\,,
\qquad \hat Y'_{ij} = X_{ij} 
+ \nabla_i \nabla_j Z
     - g_{ij}\, \square Z\,,
\ee  
after imposing the (Minkowski)$_3\times M_8$ background equations. The
variation of the additional $D=8$ Euler integrand term $2\hat Y_2\sqrt{-\hat
g}$ yields a contribution $-\hat g_{\mu\nu}\, \hat Y_2$ in the 3
spacetime directions, and zero in the internal directions (since $\hat Y_2$ 
is topological in eight dimensions).

   The variation of the full $\hat W$ term in the M-theory effective 
action therefore leads to the corrected Einstein equations
\bea
\hat R_{\mu\nu} -\ft12 \hat R\, \hat g_{\mu\nu} &=& 
 -\fft{\beta}{1152}\, (\square Z + Y_2)\, g_{\mu\nu}
\,,\label{einstmunu}\\
\hat R_{ij} - \ft12 \hat R\, \hat g_{ij} &=& \fft{\beta}{1152}\, 
(X_{ij} +\nabla_i \nabla_j\, Z - g_{ij}\, \square Z)\,,\label{einstij}
\eea
after imposing the (Minkowski)$_3\times M_8$ structure in the $\beta$
correction terms.  We do not need to include the energy-momentum
tensor of the 4-form here, since $\hat F_\4$ is taken to vanish at
leading order, and thus it itself will be of order $\beta$ in the
corrected solutions and so it would contribute only at order $\beta^2$
in the Einstein equations. For the same reason, we do not need to
include the contribution to the Einstein equation that would come from
varying the metrics in the $\hat A_3\wedge \hat X_8$ term in
(\ref{mmod2}), since it already carries a factor of $\beta$, and since
the resulting $\hat F_4$ will also be small, of order $\beta$.

    The corrected field equation for $\hat F_\4$ is
\be
d{\hat * \hat F_\4} = \ft12 \hat F_\4\wedge \hat F_\4 + 
(2\pi)^4\, \beta\, \hat X_\8\,.\label{f4inhom}
\ee
The 4-form and the eleven-dimensional metric will be required to have
the 3-dimensional Poincar\'e invariance of the leading-order solution,
which implies that we can write
\bea
d\hat s_{11}^2 &=& e^{2A}\, \eta_{\mu\nu}\, dx^\mu\, dx^\nu + e^{-A}\,
     ds_8^2\,,\label{11met}\\
\hat F_4 &=& d^3x \wedge df + G_\4\,,\label{11f}
\eea
where $A$ and $f$ are functions only of the coordinates on $M_8$, and $G_\4$ 
is a 4-form residing purely in the internal space.  

\subsection{Spin(7) non-compact solutions}

The discussion that follows will be similar to one given in 
Ref.\ \cite{becbec2}. 
Since we are working only to order $\beta$ in this discussion, we can 
consider separately the
contributions of the two terms in the field-strength ansatz (\ref{11f}).  
The former is obligatory, in the sense that the local equation of motion
(\ref{f4inhom}) forces $f$ to become non-zero (and of order $\beta$). 
In contrast, the inclusion of the second term $G_\4$ in (\ref{11f}) is optional
if the ``internal'' space $K_8$ is non-compact; in particular it can
be chosen to be zero. To proceed, we consider this case first,
subsequently returning to consider the modifications needed for
compact $K_8$. 

The Ricci tensor of the metric (\ref{11met}) has non-vanishing 
coordinate-frame components given by
\bea
\hat R_{\mu\nu} &=& - e^{3A}\, \square A \, \eta_{\mu\nu}\,,\label{ricsmu}\\
\hat R_{ij} &=& R_{ij} + \ft12\, \square A\, g_{ij} - \ft92 \nabla_i A\, 
  \nabla_j A\,,\label{rics}
\eea
where $R_{ij}$ is the Ricci tensor of $ds_8^2= g_{ij}\, dy^i\, dy^j$.
Note that since we shall be working to order $\beta$, and since the
leading-order background is $d\hat s_{11}^2 = \eta_{\mu\nu}\, dx^\mu\, 
dx^\nu + ds_8^2$ where $ds_8^2$ is Ricci-flat, we may neglect the
terms quadratic in $\nabla A$ in the expression for $\hat R_{ij}$, 
since we shall have
\be
A = 0 + {\cal O}(\beta)\,.
\ee
Similarly, exponential factors of $e^A$ that multiply quantities that
are already of order $\beta$ may be replaced by $1$.  We shall drop
all such higher-order terms in what follows.  In particular, we may
write (\ref{rics}) simply as
\bea
\hat R_{\mu\nu} &=& - \square A \, \eta_{\mu\nu}\,,\label{rics2munu}\\
\hat R_{ij} &=& R_{ij} + \ft12 \, \square A\, g_{ij}\,.\label{rics2ij}
\eea

    From (\ref{rics2munu}) and (\ref{rics2ij}) we find $\hat R= R +
\square A$, and hence by substituting (\ref{rics2ij}) into
(\ref{einstij}) we find
\be
R_{ij} - \ft12 R\, g_{ij} = \fft{\beta}{1152}\, (X_{ij} + \nabla_i\nabla_j\, 
Z - g_{ij}\, \square Z)\,.\label{rij3}
\ee
Taking the trace gives $R= (\beta/576)\, \square Z$, and hence (\ref{rij3})
yields
\be
R_{ij} =  \fft{\beta}{1152}\, (X_{ij} + \nabla_i\nabla_j\, Z )\,.
\label{8comps}
\ee
 From (\ref{einstmunu}) we then find 
\be
\square A = \fft{\beta}{1728}\, Y_2\,.\label{3comps}
\ee

   Equations (\ref{8comps}) and (\ref{3comps}) comprise the final
expressions that follow from the corrected Einstein equations
(\ref{einstmunu}) and (\ref{einstij}).  It is important to note that
all terms involving $\square Z$ have cancelled.\footnote{The 
analogous cancellation did not occur in the discussion presented
in \cite{becbec2} for deformations of eight-dimensional 
Ricci-flat K\"ahler backgrounds, but this is simply because a 
different choice of  field variables was used
there.  Earlier papers, including \cite{becbec,hawtay,bricvenaq}, did not
include the contributions from the volume terms $\hat Y$ and $\hat
Y_2$ in (\ref{mmod2}) at all, and so the ``M2-brane like'' metric
ansatz (\ref{11met}) that was made in those papers would have been in
conflict with the Einstein equations in the spacetime directions at
order $\beta$ (see (\ref{ricsmu}), (\ref{rics}), (\ref{rics2munu}) and
(\ref{rics2ij})).}  This depends, in particular, on the fact that
$X_{ij}\, g^{ij}=\square Z$, which was shown for a Spin(7) background
in (\ref{trxij}).  Note that the correction to the Ricci-flatness of the
leading-order Spin(7) manifold, described by (\ref{8comps}), is
identical to the corrected equation (\ref{einstmod}) that we obtained
at tree level in string theory.

   Again working to order $\beta$, substitution of the ansatz 
(\ref{11f}) into the corrected 4-form equation (\ref{f4inhom}) yields
$d{*d}f= \beta\, (2\pi)^4\, X_8$, or, after dualization
\be
\square \, f = \beta \, (2\pi)^4\, {*X_8}\,.\label{feqn}
\ee

    If the internal space $M_8$ admits a nowhere-vanishing spinor, as
is always the case on a space of special holonomy, there is a
topological relation between the Euler class $E_8$ and the
combination of $P_2$ and $P_1^2$ Pontryagin classes that arises in
$X_8$ \cite{ishpop1,ishpop2}.  This translates into the statement that
\be\label{Y2identity}
Y_2= 576 (2\pi)^4\, {*X_8}\,.
\ee
Comparing (\ref{3comps}) and (\ref{feqn}), this implies (for non-singular
solutions without $\delta$-function sources) that we must have
\be
f = 3A\,.\label{farel}
\ee
As we shall now show, this is in fact precisely the condition that is
needed in order to ensure that the deformed solution will still be
supersymmetric.

\subsection{Supersymmetry of the deformed 
             (Minkowski)$_3\times K_8$ background}

   The classical gravitino transformation rule in eleven-dimensional
supergravity takes the form 
\be
\delta \hat \psi_{\sst M} = \hat\nabla_{\sst M}\, \hat\epsilon - \ft1{288}
\hat F_{\sst{N_1\cdots N_4}}\,
\hat\Gamma_{\sst M}{}^{\sst{N_1\cdots N_4}} \,\hat\epsilon
  +\ft1{36} \hat F_{\sst{M N_1\cdots N_3}}\, 
\hat \Gamma^{\sst{N_1\cdots N_3}}\, \hat\epsilon \,.\label{classsusy}
\ee
We shall use the following $11=3+8$ decomposition of the eleven-dimensional 
Dirac matrices $\hat\Gamma_M$:
\be
\hat\Gamma_\mu = \gamma_\mu\otimes \Gamma_9\,,\qquad
\hat\Gamma_i = \oneone\otimes \Gamma_i\,,
\ee
where $\Gamma_9$ is the chirality operator in the eight-dimensional
internal space.  To the order $\beta$ that we are working, it suffices
to retain the contributions from the field strength $\hat F_\4$ and
the metric warp factor $A$ only up to linear order.  From
(\ref{11met}), we therefore find that in the natural choice of spinor
frame, the covariant derivative $\hat\nabla_M$ in the spacetime and
internal directions is given by
\be 
\hat\nabla_\mu = \del_\mu\otimes \oneone + \ft12 \del_i A\, \gamma_\mu\otimes
\Gamma_9\Gamma^i\,,\qquad 
\hat\nabla_i =\oneone\otimes\nabla_i - \ft14 \del_jA\, \oneone\otimes 
\Gamma_i{}^j\,.
\ee
Including the contribution of the 4-form, which is given by (\ref{11f}), 
we therefore have the supersymmetry transformation $\delta\hat\psi_M=
\hat D_M\, \hat\epsilon$, where
\bea
\hat D_\mu &=&\del_\mu - \ft12 \del_i A\, \gamma_\mu\otimes
\Gamma^i \, \Gamma_9 - \ft16 \del_i f\, \gamma_\mu\otimes\Gamma^i\,,\nn\\
\hat D_i &=& \oneone\otimes \nabla_i - \ft14 \del_j A\, \oneone\times 
\Gamma_i{}^j -\ft1{12}\del_j f\, \oneone\otimes \Gamma_i{}^j\, \Gamma_9 
   +\ft16 \del_i f\, \oneone\otimes \Gamma_9 + \oneone\otimes Q_i\,,
\label{supercov}
\eea
and $Q_i$ is the correction to the supersymmetry transformation discussed
in section \ref{susycorrectsec}.  It is straightforward to verify that
the Killing spinor condition $\hat D_M\, \hat\epsilon=0$ is satisfied if 
we write
\be
 \hat\epsilon = e^{\ft12 A}\, \epsilon\otimes\eta\,,
\ee
where $\epsilon$ is a constant spinor in the 3-dimensional Minkowski 
spacetime, and $\eta$ is a chiral spinor in the internal 8-dimensional
space, $\Gamma_9\, \eta=-\eta$, which satisfies the usual modified 
covariant-constancy condition
\be
\nabla_i\, \eta + Q_i\, \eta=0
\ee
that we discussed previously in the context of tree-level string 
corrections.

   Note that the additional ingredients in the current M-theory
discussion, in comparison to our previous tree-level string
discussion, are associated with the warp factor appearing in the
metric (\ref{11met}), and the field strength (\ref{11f}) that is
forced to be non-zero because of the $\hat A_3\wedge \hat X_8$ term in
the effective action.  These two contributions in the supercovariant
derivatives (\ref{supercov}) cancel against each other, by virtue of 
(\ref{farel}), in exactly the
same way as one finds in a standard M2-brane solution \cite{dust} of
eleven-dimensional supergravity.

\subsection{Compact $K_8$}

When the internal manifold $K_8$ is non-compact then the inclusion of
the term $G_\4$ in the field-strength ansatz (\ref{11f}) is
optional. However, when $K_8$ is a compact manifold of non-zero Euler
number there is an additional topological condition that follows by
integrating (\ref{f4inhom}), namely \cite{hawtay}
\be
\int_{K_8} G_\4\wedge G_\4 = \fft{(2\pi)^4\, \beta}{12}\, \chi\,,
\label{globalcon}
\ee
where $\chi$ is the Euler number of $K_8$.  Under these circumstances, the
inclusion of the term $G_\4$ in (\ref{11f}) becomes obligatory;
clearly we must take
\be
G_\4 = \sqrt{\beta}\, \omega_\4
\ee
where $\omega_\4$ is a closed 4-form on $K_8$ that we take to be
$\beta$-independent. It must also be co-closed in order to avoid an order
$\sqrt{\beta}$ correction in (\ref{f4inhom}). There is also a 
potential order
$\sqrt{\beta}$ correction to the supercovariant derivatives
(\ref{supercov}), namely 
\bea
\hat D_\mu &\longrightarrow & \hat D_\mu  -\sqrt{\beta} 
\ft1{288}\, \omega_{j_1\cdots j_4}\, 
\gamma_\mu \otimes \Gamma^{j_1\cdots j_4}\,,\nn\\
\hat D_i &\longrightarrow & \hat D_i -\sqrt{\beta}
\ft1{288}\, \oneone\otimes
( \omega_{j_1\cdots j_4}\, \Gamma_i{}^{j_1\cdots j_4} - 8 
\omega_{i j_1\cdots j_3}\, \Gamma^{j_1\cdots j_3})\,.
\eea
The $\sqrt{\beta}$ corrections cancel if
\be
\omega_{ij_1\cdots j_3}\, \Gamma^{j_1\cdots j_3}\, \eta=0
\label{omcon}
\ee
is satisfied.  This can be viewed as a supersymmetry-preservation 
condition on the internal 4-form $\omega_\4$. It implies that $\omega_\4$
must be self-dual \cite{becbec,hawtay,kbec} (which is the same sense of
duality as for the calibrating 4-form $c_\4$ given by (\ref{c4def})), and
hence that it must be closed as well as co-closed. In other words,
$\omega_4$ must be a self-dual harmonic 4-form. Note, however, that 
$c_\4$ is not a suitable candidate for $\omega_\4$ because
if we left-multiply (\ref{omcon}) by $\bar\eta\Gamma^i$ we get 
$\omega_{ijk\ell}\, c^{ijk\ell}=0$, and this is not satisfied by
$\omega_4=c_4$. What this shows is that $\omega_4$ must be a 
self-dual harmonic 4-form that is {\it orthogonal} to $c_\4$.  

   It is useful at this point to look at the decomposition of the
$SO(8)$ tangent-space representations of 4-forms under the Spin(7)
holonomy group.  We have
\be
{\bf 35}_+ \longrightarrow {\bf 1} + {\bf 7} + {\bf 27}\,,\qquad 
{\bf 35}_- \longrightarrow {\bf 35}\,,
\ee
for self-dual and anti-self-dual 4-forms respectively.  Since this
decomposition is made with respect to the invariant calibrating
4-form, which defines the Spin(7) embedding in $SO(8)$, it follows
that the decomposition commutes with covariant differentiation.  This
allows a refinement of the cohomology for self-dual 4-forms, 
in which we may write \cite{joy}
\be
H^4_+(K_8,\R) = H_1^4(K_8,\R) + H_7^4(K_8,\R) + H_{27}^4(K_8,\R)\,.
\ee
Correspondingly, we have for the Betti numbers $b_4=b_4^+ + b_4^-$,
with $b_4^+= b_4^{(1)} + b_4^{(7)} + b_4^{(27)}$.  It is shown in
\cite{joy} that for any compact Spin(7) manifold, 
$b_4^{(7)}=0$, and $b_4^{(1)}=1$.  This last identity corresponds to
the fact that the calibrating 4-form is the unique Spin(7)-invariant
self-dual harmonic form.  Thus we have that
\be
b_4^+ = 1 + b_4^{(27)}\,,\label{b27}
\ee
and so {\it any} self-dual harmonic 4-form other than the calibrating
4-form can provide a solution that satisfies the supersymmetry
condition (\ref{omcon}).\footnote{In fact, a more detailed investigation of
(\ref{omcon}) reveals that it already selects precisely self-dual
4-forms in the {\bf 27} representation of Spin(7), quite independently of
the above discussion of the refined cohomology.}

    The fact that $\omega_\4$ is closed takes care of any  order
$\sqrt{\beta}$ terms in (\ref{f4inhom}), but we must now take into account the
order $\beta$ contribution from the $\hat F_4\wedge \hat F_4$
term. This has the effect of modifying (\ref{feqn}) to
\be
\square \, f = \beta \, [(2\pi)^4\, {*X_8} +\ft1{48} |\omega_\4|^2]\,,
\label{mod.feqn}
\ee
where we have used the self-duality of $\omega_\4$ to write the dual of
$\omega_\4\wedge \omega_\4$ as $\ft1{24}|\omega_\4|^2$. There is a similar 
order $\beta$ correction to the
stress-tensor for $\hat F_4$ (which we were previously able to set to
zero). This modifies the source for the Einstein equations on $K_8$, 
but the only effect of this is a modification of the
source term of the Poisson equation (\ref{3comps}), which becomes
\be
\square A = \beta\, [\ft{1}{1728}\, Y_2 + \ft{1}{144}\, |\omega_\4|^2]\,.
\label{mod.3comps}
\ee
Fortunately, the consistency of (\ref{mod.feqn}) and
(\ref{mod.3comps}) is again assured because of  
(\ref{Y2identity}), and again we find that
$f=3A$, just as we found for a non-compact $K_8$ with vanishing $G_4$.
As we saw in the non-compact case, the equality $f=3A$ is crucial for
the supersymmetry of the deformed background. Note that this 
is also the relation found in Ref \cite{greenvanhove} from direct 
consideration of supersymmetry in the $R^4$ -- Chern Simons system. 

   To summarise, we have shown that if $G_\4$ is taken to be
proportional to {\it any} self-dual harmonic 4-form other than the
calibrating 4-form (\ie any self-dual harmonic 4-form in the {\bf 27} of
Spin(7)), the local equations of motion and the supersymmetry
conditions are still satisfied by the deformed Spin(7) holonomy background,
up to order $\beta$.  Furthermore, one can always satisfy the global
topological constraint (\ref{globalcon}), by normalising the 
harmonic self-dual 4-form appropriately, namely so that
\be
\int_{K_8} |G_\4|^2 = 2(2\pi)^4\, \beta\, \chi\,,
\label{globalcon2}
\ee
In \cite{joy}, many examples of compact manifolds with Spin(7)
holonomy are constructed, typically with large values of the Betti
number $b_4^+$ of self-dual harmonic 4-forms.  In fact from
(\ref{b27}), we see that whenever $b_4^+$ is greater than 1, there
will exist suitable self-dual harmonic 4-forms that allow the global
condition (\ref{globalcon}) to be satisfied.  

    It is also worth noting that if $K_8$ is non-compact, in which
case the inclusion of a non-vanishing $G_\4$ is optional rather than 
obligatory, explicit constructions of self-dual harmonic forms 
that satisfy the supersymmetry condition (\ref{omcon}) are known 
\cite{gibpagpop,clptrans,newspin7}.

\section{Deformation of $SU(5)$ Holonomy Solutions of M-theory}

We now turn to compactifications of M-theory on ten-dimensional
manifolds $K_{10}$ which, at leading order, are Ricci-flat and
K\"ahler.  It should be emphasised that such backgrounds probe aspects
of M-theory that go beyond anything that can be directly deduced from
light-cone string-theory computations, which, in practice, have
provided most of the concrete information about the structure of
M-theory.  In fact, $SU(5)$ holonomy backgrounds cannot be discussed
at all in perturbative string theory, since there are only nine
Euclidean-signature dimensions.  Thus not only do $SU(5)$ holonomy
backgrounds go beyond what can be learned from light-cone
string-theory calculations, they go beyond perturbative string theory
itself, and are intrinsic to M-theory.  Nevertheless, it has been
argued that the information learned from light-cone string
calculations, and elsewhere, can be extrapolated to genuinely
eleven-dimensional results about the structure of M-theory.  It is
therefore of interest to see what happens if one tries to ``push the
envelope'' and apply these eleven-dimensional results to $SU(5)$
holonomy backgrounds.

\subsection{Leading-order preliminaries}
\label{sect.prelim} 

To set up our discussion of corrections to $SU(5)$-holonomy
compactifications of M-theory, we will begin with a brief discussion
of the leading-order $SU(5)$-holonomy compactifications of
11-dimensional supergravity. The (undeformed) solutions of interest
have vanishing fermions, vanishing 4-form field strength $F$, and a
metric of the form
\be
ds^2 \equiv g_{MN}dx^Mdx^N =  -dt^2 + g_{ij}dx^idx^j
\ee
where the 10-metric $g_{ij}$ on $K_{10}$ has $SU(5)$ holonomy. The 11D
Dirac matrices can be taken to be
\be
\hat\Gamma_0 = i\gamma_{11}, \qquad \hat \Gamma_i = \gamma_i
\ee
where $\gamma_i$ are the $SO(10)$ Dirac matrices, and $\gamma_{11}$ is
the chirality operator on $SO(10)$ spinors,
\be
\gamma_{11} = i\gamma_1\gamma_2\cdots\gamma_{10}. 
\ee
We will assume (in accordance with the usual custom) that the 11D
Dirac matrices $\hat\Gamma_M$ are hermitian, in which case the
$SO(10)$ Dirac matrices $\gamma_i$ are hermitian.

The supersymmetry-preservation condition for solutions of 11D
supergravity is the vanishing of the supersymmetry variation of the
gravitino. For purely gravitational backgrounds this reduces to
\be
\label{delta.gravitino}
\hat D_M\hat\epsilon =0
\ee
where $D_M$ is the covariant derivative on spinors and $\hat\epsilon$
is a Majorana spinor; i.e., it satisfies
\be\label{Maj11}
\hat\epsilon^\dagger  = \hat\epsilon^T \hat C\hat \Gamma_0
\ee
where $\hat C$ is the antisymmetric $SO(1,10)$ charge conjugation
matrix.  For compactifications on $K_{10}$, the condition
(\ref{delta.gravitino}) reduces to the equation
\be\label{susypres.lowest}
D_i\hat\epsilon =0, 
\ee
where $\hat\epsilon$ is now a time-independent $SO(10)$ spinor on
$K_{10}$ and $D_i$ is the covariant derivative on such spinors. The
11D Majorana condition (\ref{Maj11}) becomes
\be
\hat\epsilon^* = C\hat\epsilon\, ,\qquad C= \hat C\hat \Gamma_0
\ee
where $C$ is the real symmetric $SO(10)$ charge conjugation matrix,
with the property that
\be
\qquad C\gamma_i C^{-1} = \gamma_i^T\, .
\ee
Equivalently, since the matrices $\gamma_i$ are hermitian,
\be
C\gamma_i C^{-1} = \gamma_i^*\, .
\ee
One also has $C^2=1$, so one may choose a basis such that $C=1$, in
which case the matrices $\gamma_i$ are real, as are Majorana
spinors. However, in this basis $\gamma_{11}$ is pure imaginary, so
the Majorana condition is not compatible with a chirality
condition. This result is, of course, basis independent, so a
`minimal' $SO(10)$ spinor is {\it either} Majorana {\it or} complex
chiral.

For some purposes it is simpler to work with complex chiral $SO(10)$
spinors. In particular, a 10-manifold of $SU(5)$ holonomy admits one
covariantly constant complex chiral spinor, as follows from the
decomposition
\be
{\bf 16} = {\bf 10} \oplus  {\bf 5} \oplus {\bf 1}
\ee
of the spinor irrep of $Spin(10)$ into irreps of $SU(5)$. Let $\eta$
be this one chiral spinor; we choose conventions such that the
chirality condition is
\be\label{chirality}
\gamma_{11} \eta = -\eta\, .
\ee
Note that the charge conjugate spinor 
\be
 \eta^c := C^{-1}\eta^*
 \ee
satisfies the anti-chirality condition $\gamma_{11}\eta^c = \eta^c$ as
a consequence of the identity (for hermitian Dirac matrices)
 \be
 C\gamma_{11}C^{-1} = -\gamma_{11}^*. 
 \ee
Moreover, as a consequence of the identity
\be
 C\gamma_{ij}C^{-1} = \gamma_{ij}^*,
 \ee
the spinor $\eta^c$ is covariantly constant if $\eta$ is covariantly 
constant. An alternative way to see this is to note that the covariant 
derivative is real in a real basis for the Dirac matrices, so that 
in such a basis the real and imaginary parts of a covariantly constant 
complex spinor are covariantly constant Majorana spinors. In particular, the 
existence of one covariantly constant chiral spinor $\eta$ implies the
existence of two linearly independent covariantly constant Majorana
spinors, defined by
 \be
\epsilon_1 = {1\over2} \left( \eta + \eta^c\right)\, ,\qquad
\epsilon_2 = -{i\over2}\left(\eta - \eta^c\right).
\ee
Using $C^2=1$, it is easily verified that these spinors {\it are} 
Majorana. They are covariantly constant because $D_i\eta=0$
implies $D_i\eta^c=0$. Note that
\be
\eta = \epsilon_1 + i\epsilon_2\, 
\ee
which is the decomposition of a complex spinor into two Majorana
spinors; in a real basis, for which $C=1$, the Majorana spinors
$\epsilon_1$ and $\epsilon_2$ are just the real and imaginary parts of
the complex spinor $\eta$. If $\eta$ is a chiral spinor, satisfying
(\ref{chirality}) then
\be
 \epsilon_2 = i\gamma_{11}\epsilon_1\, .
\ee
However, the Majorana spinors $\epsilon_1$ and $\epsilon_2$ are still
linearly independent over the reals because the linear combination $(a
+ib\gamma_{11})\epsilon_1$ vanishes for real numbers $a,b$, and
non-zero $\epsilon_1$, if and only if $a=b=0$.

    We have thus shown that, when $K_{10}$ is a manifold of $SU(5)$
holonomy, there are two linearly-independent Majorana spinor solutions
of (\ref{susypres.lowest}), and hence of the supersymmetry
preservation condition (\ref{delta.gravitino}), and that this
statement is equivalent to the statement that $K_{10}$ admits a single
complex chiral Killing spinor.
 
 For future use we also note that
\be
 \gamma_{\hat j}\,\eta = \im\, 
\gamma_j\, \eta \,,\qquad (\gamma_{ij} + \gamma_{\hat i\hat j})\,
\eta = 2\im\, J_{ij}\, \eta\,,
\ee
where $J_{ij}$ is the K\"ahler form, and we are using the ``hat''
notation of \cite{fpss}, defined in (\ref{hatdef}). Other useful
properties following from these are
\be
\bar\eta\, \gamma_{ij}\, \eta = \im\, J_{ij}\,,\qquad
\bar\eta\, \gamma_{ijk\ell}\,\eta = -J_{ij}\, J_{k\ell} -
J_{ik}\, J_{\ell j} - J_{i\ell}\, J_{jk}\,.
\ee

\subsection{Corrections to  (Minkowski)$_1\times K_{10}$ backgrounds}
\label{k10sec1}

   The relevant ${\cal O}(\beta)$ corrections to the equations of
motion again follow from (\ref{mmod2}).  The contributions from the
eight-dimensional Euler integrand term $\hat Y_2\, \sqrt{-\hat g}$ can
be determined by varying the explicit metrics needed to write
$\sqrt{-\hat g}$ times the right-hand side of (\ref{y2exp}) in terms
of canonical Riemann tensors $\hat R^M{}_{NPQ}$ with one index up and three
down.  (One does not need to vary the metrics from which $\hat R^M{}_{NPQ}$
is constructed, since these variation terms will be of the form of a total
derivative, and hence will not contribute in the equations of
motion.\footnote{In the same way, the terms from the metrics in
$R_{MN}=R^P{}_{MPN}$ do not contribute when one varies the
two-dimensional Euler integrand $g^{MN}\, R_{MN}\, \sqrt{-g}$ (the
Einstein-Hilbert action) to obtain the Einstein tensor.})  Thus defining
$\delta \int \hat Y_2\, \sqrt{-\hat g}= \int \sqrt{-\hat g}\, 
\hat E_{\sst{MN}}\, \delta \hat g^{\sst{MN}}$, one finds (see, for example, 
\cite{derubust})
\be
\hat E_{\sst{M}}{}^{\sst{N}} 
= - \fft{9!}{2^9}\, \delta_{\sst{M M_1\cdots M_8}}^{\sst{N N_1\cdots N_8}}\,
    \hat R^{\sst{M_1 M_2}}{}_{\sst{N_1 N_2}}\cdots 
\hat R^{\sst{M_7 M_8}}{}_{\sst{N_7 N_8}}\,,
\label{eulervar}
\ee
where the Kronecker deltas are of unit strength 
($\delta_{\sst{M_1 \cdots M_n}}^
{\sst{N_1 \cdots N_n}} \omega_{\sst{N_1\cdots N_n}}=
\omega_{\sst{M_1\cdots M_n}}$ for any
antisymmetric tensor $\omega_{\sst{M_1\cdots M_n}}$).
  
    The eleven-dimensional Einstein equations, with their ${\cal O}(\beta)$ 
corrections, are given by
\bea
\hat R_{00} -\ft12 \hat R\, \hat g_{00} &=& 
 -\fft{\beta}{1152}\, \square Z \, g_{00} + \fft{\beta}{576}\, \hat E_{00}
\,,\label{einst10munu}\\
\hat R_{ij} - \ft12 \hat R\, \hat g_{ij} &=& \fft{\beta}{1152}\, 
(X_{ij} +\nabla_i \nabla_j\, Z - g_{ij}\, \square Z)
+ \fft{\beta}{576} \, \hat E_{ij}
\,,\label{einst10ij}
\eea
where
\be
Z= R_{ijk\ell}\, R^{k\ell mn}\, R_{mn}{}^{ij} - 2
   R_{ikj\ell}\, R^{k m\ell n}\, R_m{}^i{}_n{}^j\,,\label{zzdef}
\ee
after imposing the (Minkowski)$_1\times K_{10}$ Ricci-flat K\"ahler
background conditions in the correction terms on the right-hand sides.
Note that we shall have
\bea
\hat E_{00} &=& \ft12 Y_2\,,\nn\\
\hat E_i{}^j &=& E_i{}^j \equiv - \fft{9!}{2^9}\, 
\delta_{i i_1\cdots i_8}^{j j_1\cdots j_8}\,
    R^{i_1 i_2}{}_{j_1 j_2}\cdots 
     R^{i_7 i_8}{}_{j_7 j_8}\,,\label{euler8var2}
\eea
in the (Minkowski)$_1\times K_{10}$ background.  The new feature that
we encounter here, in comparison to the (Minkowski)$_3\times K_8$
backgrounds described by (\ref{einstmunu}) and (\ref{einstij}), is
that in (\ref{einst10ij}) we have the non-zero contribution $\hat
E_{ij}$ coming from the variation of the eight-dimensional Euler
integrand.  It is manifest from its form, given in (\ref{euler8var2}),
that this would vanish in an 8-dimensional curved background, owing to
the antisymmetrisation over 9 indices.

   As in the case of (Minkowski)$_3\times K_8$ backgrounds, we expect
that the effect of the order $\beta$ corrections to the
(Minkowski)$_1\times K_{10}$ background will be to introduce a warp
factor in the eleven-dimensional metric, as well as causing the
originally-vanishing 4-form to become non-zero.  For the metric, we
therefore write
\be
d\hat s_{11}^2 = -e^{2A}\,dt^2 + e^{-\ft14 A}\, ds_{10}^2\,,\label{d10warp}
\ee
where the function $A$ in the warp factor depends only on the
coordinates of $K_{10}$.  The relative powers of the warp factor in
the two terms in (\ref{d10warp}) are motivated by the expectation of a
``0-brane'' structure in the deformed solution.  At the linearised
level, which suffices for our purposes since we are perturbing around
the original background with $A=0$ and $K_{10}$ Ricci-flat and
K\"ahler, we find that the non-vanishing Riemann tensor components for
the metric (\ref{d10warp}) are given by
\bea
\hat R_{0i0j} &=& \nabla_i\nabla_j A\,,\\
\hat R_{ijk\ell} &=& R_{ijk\ell} - \ft18(g_{i\ell}\, \nabla_j\nabla_k A
- g_{ik}\, \nabla_j\nabla_\ell A +  g_{jk}\, \nabla_i\nabla_\ell A
-g_{j\ell}\, \nabla_i\nabla_k A)\,,
\eea
and the non-vanishing components of the Ricci tensor are given by
\be
\hat R_{00} = \square A\,,\qquad \hat R_{ij} = R_{ij} + \ft18 g_{ij}\, 
\square A\,.\label{1110eq}
\ee
Taking the eleven-dimensional trace gives $\hat R= R + \ft14 \square A$,
and substituting this into (\ref{einst10ij}) and tracing leads to
\be
R = \fft{\beta}{576}\, \square Z - \fft{\beta}{2304}\, E_i{}^i\,.
\label{tracer}
\ee
Equation (\ref{einst10ij}) then gives 
\be
R_{ij} = \fft{\beta}{1152}\, \Big( X_{ij} + \nabla_i\nabla_j\, Z 
    + 2E_{ij} - \ft14 E_k{}^k\, g_{ij}\Big)\,.\label{rijeq10}
\ee
Note that $X_{ij}$, coming from the variation of the ``string tree-level''
term $\hat Y$, is given by $X_{ij} = \nabla_{\hat i} \nabla_{\hat j}\, Z\equiv
J_i{}^k\, J_j{}^\ell \,\nabla_k
\nabla_\ell\, Z$, as usual in a K\"ahler background.  Note also that
from (\ref{eulervar}) we shall have 
\be
E_k{}^k = - Y_2\,.\label{tracee}
\ee
The remaining content of the Einstein equations is contained in 
(\ref{einst10munu}).  From (\ref{tracer}) and (\ref{tracee}), we find
that this implies
\be
\square A = \fft{\beta}{1728}\, Y_2\,.\label{sqa}
\ee
After using (\ref{tracee}), equation (\ref{rijeq10}) can be written as
\be
R_{ij} = \fft{\beta}{1152}\, \Big( \nabla_{\hat i} \nabla_{\hat j}\, Z 
 + \nabla_i\nabla_j\, Z 
    + 2 E_{ij} + \ft14 Y_2\, g_{ij}\Big)\,.\label{rijeq102}
\ee

    Equations (\ref{sqa}) and (\ref{rijeq102}) determine the warp factor
and the Ricci tensor of the corrected ten-dimensional K\"ahler metric,
respectively. The field equation (\ref{f4inhom}) will govern
the structure of the non-vanishing 4-form that is required at order $\beta$.
In order to maintain the 1-dimensional ``Poincar\'e symmetry'' of the
original uncorrected background, it must be that
\be
\hat F_\4 = G_\3\wedge dt + G_\4\,,\label{d104form}
\ee
where $G_\3$ and $G_\4$ are 3-form and 4-form fields on $K_{10}$.  We
may, to begin with, assume that $G_\4=0$.  The 4-form equation of
motion (\ref{f4inhom}) then implies, up to order $\beta$, that we shall have
\be
d{*G_\3}= (2\pi)^4\, \beta\, X_8\,,\label{g3eqn}
\ee
where the unhatted $*$ denotes Hodge dualization in $K_{10}$.  

Since the integrability condition obtained by taking the exterior
derivative of this equation is trivially satisfied, we are guaranteed
to be able to find a local solution of (\ref{g3eqn}). However,
integration over any 8-cycle $C_8$ of $K_{10}$ leads to \be \int_{C_8}
X_8 =0\, , \ee which must be satisfied for {\it all} 8-cycles
$C_8$. This is a topological constraint on $K_{10}$ that will not in
general be satisfied unless $H_8(K_{10})$ is trivial. As Poincar\'e
duality implies that $H_8 \cong H_2$ for any compact 10-manifold, and
as $H_2$ is necessarily non-trivial for any K\"ahler manifold, the
topological constraint is not satisfied by any compact K\"ahler
10-manifold; in other words, it is not satisfied by any compact
manifold $K_{10}$ of $SU(5)$-holonomy. What this means is that it is
inconsistent to set $G_{(4)}$ to zero (as we have been doing) when
$K_{10}$ is compact.  In principle, we could attempt to take this into
account as we did in the Spin(7) case by allowing for a non-zero
$G_{(4)}$ of order $\sqrt{\beta}$.  However, the implications for
supersymmetry are much less straightforward than they were for Spin(7)
compactifications, so we shall not attempt an analysis along these
lines here.  Instead, we shall simply restrict discussion to the class
of $SU(5)$-holonomy manifolds $K_{10}$ for which $H_8$ is
trivial. This implies that $K_{10}$ is non-compact, so we are
restricted to a special class of $SU(5)$-holonomy
`non-compactifications'.

With this restriction understood, the results above show 
that we can obtain an M-theory corrected
solution, at order $\beta$, to the original (Minkowski)$_1\times
K_{10}$ vacuum of $D=11$ supergravity.  The corrected metric is of
the form of a warped product (\ref{d10warp}), with the warp factor
given by (\ref{sqa}), and the Ricci tensor of $K_{10}$ given by
(\ref{rijeq102}).   In the next subsection, we shall analyse the 
question of whether
this M-theory corrected solution preserves the supersymmetry of the
original Ricci-flat K\"ahler solution of $D=11$ supergravity.

\subsection{Supersymmetry of the deformed
            (Minkowski)$_1\times K_{10}$ backgrounds}

   We have seen in the previous subsection that the Ricci tensor of
the originally Ricci-flat ten-dimensional K\"ahler space $K_{10}$ suffers
a more substantial deformation than has been seen hitherto for 
spaces $K_n$ of special holonomy with $n\le 8$, on account of the
$E_{ij}$ and $Y_2\, g_{ij}$ terms in (\ref{rijeq102}) that come from 
the variation of the Euler integrand $\hat Y_2$.  

   It is of interest now to study the supersymmetry of the corrected
(Minkowski)$_1\times K_{10}$ backgrounds.  Here, we are on somewhat
less solid ground.  Although there has been a lot of work on the
detailed structure of the higher-order corrections to supergravities
in ten and eleven dimensions (see, for example, \cite{pvhw}), there
are not, as far as we are aware, complete and explicit results for the
corrections to the supersymmetry transformation rules at order $\alp$
(or order $\beta$).  The only explicit results are those introduced in
\cite{cfpss} in the context of corrections to six-dimensional
Calabi-Yau compactifications, their extension in \cite{g2mod} to
$G_2$-holonomy compactifications, and their extension in the present
paper to Spin(7)-holonomy compactifications.  These corrections were
deduced on the basis of {\it requiring} that the unbroken
supersymmetry of the leading-order background should persist in the
face of the $\alp$ corrections.\footnote{This might seem somewhat
circular as an argument for demonstrating that supersymmetry is
preserved in the corrected special-holonomy backgrounds.  However, the
fact that one is able at all to find a candidate fully-Riemannian
correction to the gravitino transformation rule that is consistent
with the preservation of supersymmetry of the corrected backgrounds is
already quite remarkable.  And since no other explicit results for the
gravitino transformation rules have been obtained by 
direct calculation in the intervening 18 years since \cite{cfpss} 
appeared, we are forced, {\it faute de mieux}, to make do with this
at present.}
Remarkably, the same Riemannian expression (\ref{riem1})
that was first proposed in \cite{cfpss} in the six-dimensional Calabi-Yau
context has turned out to be sufficient to achieve a preservation of
supersymmetry for the $G_2$ holonomy and Spin(7) holonomy backgrounds.

For an $SU(5)$-holonomy supergravity solution of 11D supergravity, we
would again expect the M-theory correction to the gravitino
transformation rule to lead to a modified covariant derivative
$(\nabla_i + Q_i)$, where $Q_i$ is of order $\beta$. If we assume that
$Q_i$ takes the same purely Riemannian form\footnote{Note that with
the correction (\ref{riem1}) the modified Killing spinor operator
$(\nabla_i + Q_i)$ retains the same reality properties as at the
classical level, so the equivalence between a pair of Majorana spinors
and a complex chiral spinor as explained in subsection
\ref{sect.prelim} persists in the presence of the corrections.}  as in
(\ref{riem1}) then, using properties of $SU(5)$ holonomy manifolds,
one can show that
\be
Q_i = \fft{\im\, \beta}{2304}\, \nabla_{\hat i} Z\,,\label{q10}
\ee
where $Z$ is given by (\ref{zzdef}). There is no {\it a priori} reason
why this assumption should be correct; there could be further terms
whose presence would not be probed if one looked only at
(Minkowski)$_3\times K_8$ backgrounds, but which would be relevant to
(Minkowski)$_1\times K_{10}$ backgrounds.  However, we shall show that
this assumption nonetheless leads to the conclusion that supersymmetry of the
corrected $SU(5)$ holonomy backgounds is maintained, despite the loss
of $SU(5)$ holonomy.  This is {\it a posteriori} evidence that the
assumption is correct since one would hardly expect this conclusion to
follow from an incorrect assumption, irrespective of whether
supersymmetry is in fact preserved.

We begin by considering the integrability condition for the existence
of a Killing spinor that satisfies $\hat D_M\hat\ep=0$, obtained from
the commutator of supercovariant derivatives.  Since we are working
only to linear order in $\beta$, and since the field strength $\hat
F_\4$ vanishes at zeroth order, becoming non-vanishing only at order
$\beta$, we can omit terms quadratic in $\hat F_\4$ in our discussion.
We shall also suppress for now the ${\cal O}(\beta)$ $Q_i$ correction
to the supercovariant derivative; in other words, for now we shall
just consider the ``classical'' terms in the integrability condition
of $D=11$ supergravity, with the added simplification of omitting the
terms quadratic in $\hat F_\4$.  The contribution from $Q_i$ will be
included later, when we present our results.  We therefore have for
now that
\be
[\hat D_{\sst M}, \hat D_{\sst N}]_0  =  \ft14 \hat R_{\sst{MNPQ}}\, 
   \hat\Gamma^{\sst{PQ}} + \ft1{144} 
\hat\Gamma_{\sst{[M}}{}^{\sst{P_1\cdots P_4}}\, \hat\nabla_{\sst{N]}}
\, \hat F_{\sst{P_1\cdots P_4}}  
+ \ft1{18}\, \hat\nabla_{\sst{ [M}}\, \hat F_{\sst{N]P_1 P_2 P_3}}\,
\hat\Gamma^{\sst{P_1 P_2 P_3}}\,,\label{com0}
\ee
where the subscript ``0'' on the commutator indicates the omission of
the $Q_i$ correction term.  

   It is helpful to analyse the integrability conditions in stages.
First, we may note that upon left-multiplication and contraction with
$\hat\Gamma^{\sst N}$, one obtains from $\hat \Gamma^{\sst N}\, [\hat
D_{\sst M}, \hat D_{\sst N}]\, \hat\epsilon=0$ a system of field
equations that can be compared with those already derived from the
variation of the action.  Thus if a Killing spinor $\hat\epsilon$
exists, one should find consistency between the already-established
bosonic equations of motion, and those that follow from $\hat
\Gamma^{\sst N}\, [\hat D_{\sst M}, \hat D_{\sst N}]\,
\hat\epsilon=0$.  Establishing this consistency does not of itself
prove that a Killing spinor $\hat\epsilon$ exists (and thus that the
deformed solution is supersymmetric), since the left-multiplication of
the integrability condition by $\hat \Gamma^{\sst N}$ projects into a
subset of the full content of $[\hat D_{\sst M}, \hat D_{\sst N}]\,
\hat\epsilon=0$, but it already provides a non-trivial check.

   It is easy to see from (\ref{com0}) that we shall have
\be
\hat\Gamma^{\sst N}\, [\hat D_{\sst M},\hat D_{\sst N}]_0 =
-\ft12 \hat R_{\sst{MN}}\, \hat\Gamma^{\sst N} - \ft1{72}
\hat\Gamma_{\sst M}{}^{\sst{N_1\cdots N_4}}\, \hat\nabla^{\sst{N_1}}\, 
\hat F_{\sst{N_1\cdots N_4}} + \ft1{12} \hat\nabla^{\sst N}\, 
\hat F_{\sst{NMPQ}}\, \hat\Gamma^{\sst{PQ}}\,.\label{intf}
\ee
The field equation (\ref{f4inhom}) implies that
\be
\hat\nabla_{\sst M}\, \hat F^{\sst{M N_1 N_2 N_3}} =\a\, 
\hat\epsilon^{\sst{N_1 N_2 N_3 P_1\cdots P_8}}\, \hat 
X_{\sst{P_1\cdots P_8}}\,,
\label{f4eom2}
\ee
where we have, for convenience, defined
\be
\a = \fft{(2\pi)^4\, \beta}{8!}\,,
\ee
and where $\hat X_{\sst{M_1\cdots M_8}}$ denotes the components of the
8-form $\hat X_\8$, \ie
\bea
\hat X_{\sst M_1\cdots \sst M_8} &=&\fft{105}{8 (2\pi)^4}\, \Big(
R^{\sst N_1}{}_{\sst N_2 [\sst M_1 \sst M_2}\, 
R^{\sst N_2}{}_{|\sst N_3| \sst M_3 \sst M_4} 
   R^{\sst N_3}{}_{|\sst N_4| \sst M_5 \sst M_6}\, 
     R^{\sst N_4}{}_{|\sst N_1| \sst M_7\sst M_8]}\nn\\
&&\qquad \qquad  - \ft14
    R^{\sst N_1}{}_{\sst N_2 [\sst M_1 \sst M_2}\, 
       R^{\sst N_2}{}_{|\sst N_1| \sst M_3 \sst M_4} 
   R^{\sst N_3}{}_{|\sst N_4| \sst M_5 \sst M_6}\, 
      R^{\sst N_4}{}_{|\sst N_3| \sst M_7\sst M_8]}\Big)\,.\label{X88}
\eea
It is convenient also to define
\be
\hat H^{\sst{N_1 N_2 N_3}}\equiv \a\, 
\hat\epsilon^{\sst{N_1 N_2 N_3 P_1\cdots P_8}}\, \hat 
X_{\sst{P_1\cdots P_8}}\,,
\ee
so that the field equation (\ref{f4eom2}) reads
\be
\hat\nabla_{\sst M}\, \hat F^{\sst{M N_1 N_2 N_3}} = 
   \hat H^{\sst{N_1 N_2 N_3}}\,.
\ee
Since we are working only to linear order in $\beta$ (and hence $\a$), 
we are allowed to use the zeroth-order background conditions when evaluating
$\hat H^{\sst{N_1 N_2 N_3}}$.  We therefore have that the only non-vanishing
components of $\hat H_{\sst{N_1 N_2 N_3}}$ are given by
\be
\hat H_{0ij} = \a\, \ep_{ij k_1\cdots k_8}\, X^{k_1\cdots k_8}\,.
\label{hdef2}
\ee
together with those related by antisymmetry, where $\ep_{i_1\cdots i_{10}}$
is the ten-dimensional Levi-Civita tensor.

For a  K\"ahler metric on $K_{10}$, the Riemann tensor $R_{ijk\ell}$ 
satisfies
\be
R_{ijk\ell} = R_{\hat i \hat j k\ell}=R_{ij\hat k \hat \ell}\, . 
\ee
Taking into account the Riemann tensor symmetries,  this implies
that $H_{0ij}$ given in (\ref{hdef2}) will satisfy
\be 
H_{0\hat i\hat j} = H_{0ij}\,.
\ee
Taking the index value $M=0$ in (\ref{intf}) gives
\be
\hat R_{00}\, \hat \Gamma^0\, \hat\epsilon- \ft16 H_{0ij}\, 
\hat \Gamma^{ij}\,
\hat\epsilon=0\,.
\ee
Following the discussion in section \ref{sect.prelim} we may replace
the real spinor $\hat\epsilon$ by the chiral complex spinor
$\eta$. Contracting on the left with $\bar\eta$, where $\eta$ is taken
to be a Killing spinor, and using its properties as summarised in
section \ref{sect.prelim}, we deduce that $\hat R_{00} = \ft16
H_{0ij}\, J^{ij}$ and hence that
\be
\hat R_{00} = 
\ft16 \a\, J^{ij}\, \ep_{ij k_1\cdots k_8}\, X^{k_1\cdots k_8}\,.
\label{r00eqn}
\ee
Taking $M=i$ instead in (\ref{intf}), we find after some algebra
that 
\be
\hat R_{ij} = \ft1{12}\, \a\, g_{ij}\,  
J^{mn}\, \ep_{mn k_1\cdots k_8}\, X^{k_1\cdots k_8} -
\ft12 \a\, J_i{}^m\,  \ep_{jm k_1\cdots k_8}\, X^{k_1\cdots k_8}
\,.
\label{rijeqn}
\ee

   Equations (\ref{r00eqn}) and (\ref{rijeqn}) represent the gravitational
field equations that follow from the integrability conditions for the
existence of a Killing spinor.  Using (\ref{1110eq}), and now restoring the 
contribution from the $Q_i$ term in the modified supercovariant derivative, 
we  therefore find that
\bea
\square A &=& \ft16 \a\, J^{ij}\, \ep_{ij k_1\cdots k_8}\, 
X^{k_1\cdots k_8}\,,\\
R_{ij} &=&  \ft1{16}\, \a\, g_{ij}\,  
J^{mn}\, \ep_{mn k_1\cdots k_8}\, X^{k_1\cdots k_8} -
\ft12 \a\, J_i{}^m\,  \ep_{jm k_1\cdots k_8}\, X^{k_1\cdots k_8}\nn\\
&&
+\fft{\beta}{1152}\, (\nabla_{\hat i} \nabla_{\hat j}\, Z 
   + \nabla_i \nabla_j \,Z)
\,.\label{rijeq}
\eea
 From the relations between $Y_2$ and $X_8$ in a Ricci-flat K\"ahler
manifold, we can show that these equations are identical to
(\ref{sqa}) and (\ref{rijeq102}).  This establishes consistency, at least,
between the bosonic field equations and the conditions that follow
from the assumption of supersymmetry persistence in
the deformed background.  

   We now turn to consideration of the full supersymmetry
integrability conditions without taking the $\hat\Gamma^{\sst N}$
contraction; these can be read off upon substituting $\hat F_4=
G_\3\wedge dt$ into (\ref{com0}), and including also the contribution
from the $Q_i$ modification.  There are two cases to consider: taking the 
free indices $M$ and
$N$ in (\ref{com0}) to be either
$(MN)=(0i)$ or $(MN)=(ij)$.  From $(MN)=(0i)$, we find
\be
\nabla_i\nabla_j A\, \Gamma^j\, \eta = -\ft{\im}{18}\, \nabla_i 
G_{k\ell m}\, \Gamma^{k\ell m}\, \eta\,.
\ee
 From this, we find that $G_\3$ is expressible as
\be
G_\3 = \ft34 J\wedge dA + \wtd G_\3\,,\label{G3sol}
\ee
where $\wtd G_\3$ is an arbitrary 3-form that is orthogonal to the K\"ahler 
form $J$, in the sense that
\be
J^{jk}\,\wtd G_{ijk}=0\,.\label{jtrace}
\ee

   From the $(MN)=(ij)$ components of the integrability condition we find,
after substituting (\ref{G3sol}), that
\be
R_{ijk\ell}\, \Gamma^{k\ell}\, \eta + \ft{3\im}{4}\, 
(\nabla_i\nabla_{\hat j}A
-\nabla_j\nabla_{\hat i} A)\, \eta 
+ \im\, \nabla_{[i} \wtd G_{j] k\ell}\,
         \Gamma^{k\ell}\, \eta + \fft{\im\, \beta}{576}  
(\nabla_i\nabla_{\hat j}Z
-\nabla_j\nabla_{\hat i} Z)=0\,.\label{mnterms}
\ee
Multiplying by $\bar\eta$, we learn that the Ricci form $\varrho_{ij}$
is given by
\be
\varrho_{ij} \equiv \ft12 
R_{ijk\ell}\, J^{k\ell} = -\ft38 (\nabla_i\nabla_{\hat j}A
-\nabla_j\nabla_{\hat i} A)
  -\fft{\beta}{1152}  (\nabla_i\nabla_{\hat j}Z
-\nabla_j\nabla_{\hat i} Z)
\,.\label{ricciform}
\ee

    Multiplying (\ref{mnterms}) instead by $\bar\eta\, \Gamma_{mn}$,
we obtain two equations, from the real and imaginary parts.  The imaginary
part yields 
\bea
R_{ij\hat k \ell} = - R_{ijk\hat\ell} + \nabla_{[i} \wtd G_{j] k\ell} -
   \nabla_{[i} \wtd G_{j] \hat k \hat \ell}\,,\label{imag}
\eea
while the real part, after making use of (\ref{imag}), again yields 
(\ref{ricciform}).\footnote{Equation (\ref{imag}) shows that the deformed 
metric is no longer K\"ahler (at least with respect to the original
K\"ahler form $J_{ij}= -\im\, \bar\eta\Gamma_{ij}\eta$), since if it were,
the integrability condition for the covariant constancy of $J_{ij}$,
namely $[\nabla_i,\nabla_j]\, J_{k\ell}=0$, would imply that
$R_{ij\hat k \ell} = - R_{ijk\hat\ell}$.  It is perhaps useful to 
emphasise here that when looking at the Riemann tensor that arises 
from the commutation of covariant derivatives, the perturbative scheme 
in which we are working to order $\beta$ requires that we must keep terms 
of order $\beta$ that represent the deformation away from the
leading-order special-holonomy background.  By contrast, Riemann 
tensors appearing in the ${\cal O}(\beta)$ correction terms
need only be evaluated in the original undeformed special-holonomy
background.}   By making use of the cyclic
identity for the Riemann tensor, we can show from (\ref{imag}) that
\be
R_{i\hat j} = -\ft12 R_{ijk\ell} \, J^{k\ell} - \ft12\nabla^k 
\wtd G_{i\hat j\hat k} + \ft12 \nabla^k \wtd G_{ijk}\,.\label{ric1}
\ee
Note that the Bianchi identity
$d\hat F_4=0$ implies, from (\ref{G3sol}), that $d\wtd G_\3=0$, and
hence from (\ref{jtrace}) we find that $\nabla^k \wtd G_{ij\hat k}=0$, 
implying that (\ref{ric1}) reduces to 
\be
R_{i\hat j} = -\ft12 R_{ijk\ell} \, J^{k\ell} 
  + \ft12 \nabla^k \wtd G_{ijk}\,.\label{ric2}
\ee

    Substituting (\ref{ricciform}) into (\ref{ric2}), and hatting the 
$j$ index, we obtain the equation
\be
R_{ij} = \ft38 (\nabla_i\nabla_j A + \nabla_{\hat i}\nabla_{\hat j} A)
   + \fft{\beta}{1152}  
  (\nabla_i\nabla_j Z + \nabla_{\hat i}\nabla_{\hat j} Z)
- \ft12 \nabla^k \wtd G_{i\hat j k}\,.\label{ric3}
\ee
In order to verify that our assumption of supersymmetry preservation in the 
deformed system is consistent, we must show that 
(\ref{ric3}) is indeed consistent with the previous expression for
the deformed Ricci tensor as given in (\ref{rijeq10}), or, equivalently, in
(\ref{rijeq}).  This can be done by considering the equation of motion
for the 4-form field $\hat F_\4= G_\3\wedge dt$,  namely
$\nabla^k G_{ijk} = \alpha\, 
\ep_{ij \ell_1\cdots \ell_8}\, X^{\ell_1\cdots \ell_8}$. Using 
(\ref{G3sol}), this implies
\be 
\ft34 g_{ij}\, \square A - \ft34 (\nabla_i\nabla_j A +
\nabla_{\hat i}\nabla_{\hat j} A) + \nabla^k \wtd G_{i\hat j k} = 
\alpha\, \ep_{i\hat j \ell_1\cdots \ell_8}\, X^{\ell_1\cdots
  \ell_8}\,.
\ee
Substituting this into (\ref{ric3}), we obtain precisely the previous
expression (\ref{rijeq}) for the deformed Ricci tensor.

   Having verified consistency with the integrability conditions
for supersymmetry, it is instructive to examine the supercovariant
derivative itself, in the deformed $SU(5)$ holonomy background.  In
the natural orthonormal frame $\hat e^0= e^A\, dt$, $\hat e^i = 
e^{-\ft18 A}\, e^i$ for the metric (\ref{d10warp}), we find that
to linear order in the ${\cal O}(\beta)$ warp function $A$, the
torsion-free spin connection is given by
\be
\hat\omega_{0i}= -\nabla_i A\, \hat e^0\,,\qquad 
  \hat\omega_{ij} = \omega_{ij} + \ft18 (\nabla_i A\, \hat e^j-
  \nabla_j A\, \hat e^i)\,,
\ee
and hence from (\ref{classsusy}), with the correction term
(\ref{riem1}) which specialises to (\ref{q10}) in the leading-order
$SU(5)$ holonomy background, the supercovariant derivative $\hat D_A$
in the deformed background is given by
\bea
\hat D_0 &=& \del_0 - \ft{\im}{2} \nabla_i A\, \gamma^i\, \gamma_{11}
   - \ft1{36} G_{ijk}\, \gamma^{ijk}\,,\nn\\
\hat D_i &=& \nabla_i-\ft1{16}\nabla_jA\, \gamma^{ij} 
   +\ft{\im}{72} G_{jk\ell} \, \gamma_i\gamma^{jk\ell}\, \gamma_{11} 
  - \ft{\im}{8} G_{ijk}\, \gamma^{jk}\, \gamma_{11}
   +\fft{\im\, \beta}{2304} \nabla_{\hat i} Z\,,
\eea
when expressed in terms of the ten-dimensional $SO(10)$ Dirac matrices
$\gamma_i$, and the ten-dimensional chirality operator $\gamma_{11}$. 

Using these results, we find that the complex spinor $\hat\eta =
e^{\ft12 A}\, \eta$ satisfies the $D=11$ Killing spinor equation $\hat
D_A\, \hat\eta=0$ provided that $\eta$ obeys the ten-dimensional
equation
\be
D_i\eta \equiv \nabla_i\eta + \im\, (\nabla_{\hat i} h)\, \eta 
+ \ft{\im}{8} \wtd G_{ijk}\, \gamma^{jk}\eta =0\,,\label{covdir}
\ee
where
\be
h= \ft3{16} A + \fft{\beta}{2304}\, Z
\ee
together with
\be
\gamma_{11}\, \eta=-\eta\,,\qquad \wtd G_{ijk}\, \gamma^{ijk}\, \eta=0\,.
\label{gcon}
\ee
As discussed in section \ref{sect.prelim}, this result implies the
existence of two linearly-independent Majorana Killing spinors of the
M-theory background; obtained, in a real representation of the Dirac
matrices, by taking the real and imaginary parts of $\hat\eta$.

We will now show that this supersymmetry preservation by the M-theory
corrections occurs despite a deformation away from $SU(5)$ holonomy.
A straightforward calculation from (\ref{gcon}) shows that $\wtd
G_{ijk}$ is the sum of a $(1,2)$ and $(2,1)$ form, with no purely
holomorphic or anti-holomorphic $(3,0)$ or $(0,3)$ form components.
In other words,
\be
(\delta_i^\ell + \im\, J_i{}^\ell)
(\delta_j^m + \im\, J_j{}^m)
(\delta_k^n + \im\, J_k{}^n) \wtd G_{\ell mn}=0\,,
\ee
which translates, in the hatted-index notation, into the statement that
\be
\wtd G_{ijk} = \wtd G_{i\hat j\hat k} + \wtd G_{\hat i j \hat k} +
    \wtd G_{\hat i \hat j k}\,.\label{onetwo}
\ee
Using (\ref{covdir}), it is straightforward to evaluate $\nabla_j J_i{}^k$
to linear order in the deformation of the metric, where $J_{ij}= -\im\,
\bar\eta\Gamma_{ij}\eta$, yielding
\be
\nabla_j J_i{}^k = \ft12 \wtd G_{ij}{}^k - 
\ft12 \wtd G_{\hat i j}{}^{ \hat k}\,.
\label{delj}
\ee
This shows that the loss of K\"ahlerity of the leading-order $SU(5)$ 
holonomy background is associated with the non-vanishing of the
3-form $\wtd G_{ijk}$.  Calculating the Nijenhuis tensor
\be
N_{ij}{}^k = \del_{[j} J_{i]}{}^k - J_{i}{}^\ell\, 
    J_{j}{}^k\, \del_{[m} J_{\ell]}{}^k\,,
\ee
we then find from (\ref{delj}) that it is given by
\be
N_{ij}{}^k = \ft12 (\wtd G_{ij}{}^{k}  -\wtd G_{i\hat j}{}^{\hat k} - 
   \wtd G_{\hat i j}{}^{ \hat k} -\wtd G_{\hat i \hat j}{}^ {k})\,,
\ee
and so from (\ref{onetwo}) we see that the Nijenhuis tensor vanishes.
This implies that although the deformed space is no longer 
K\"ahler, it is still a complex manifold.

It is worth remarking that although the correction to the $SU(5)$
holonomy background deforms $K_{10}$ into a space that is not only
non-Ricci-flat but also non-K\"ahler, it does have the feature of 
preserving the vanishing of the first Chern class.  This can be seen
from the fact that the Ricci form, given by (\ref{ricciform}), is exact.

  \section{Conclusions}

    In this paper, we have extended the investigation of string and
M-theory corrections to special holonomy backgrounds that was begun in
Refs \cite{freemanpope,cfpss,lupost} for six-dimensional Calabi-Yau
compactifications, and subsequently developed for seven-dimensional
$G_2$ holonomy compactifications in \cite{g2mod}.  In the present
paper, we have considered the corrections at order $\alp$ in string
theory for backgrounds of the form (Minkowski)$_2\times K_8$, where
$K_8$ is a manifold of Spin(7) holonomy.  The calculations are
considerably more subtle than in the previous cases, because now there
are potential contributions to the corrected Einstein equations of a
type that would vanish identically by over-antisymmetrisation in the
case of curved backgrounds of fewer than eight dimensions.  After
handling these subtleties, we find that the corrected Einstein
equations take a rather simple form, described by (\ref{xijres}) and
(\ref{ricmod}).

   We have also considered the structure of the order $\alp$
corrections to the supersymmetry transformation rules for an
originally Spin(7) holonomy background.  Consideration of these
corrections is essential if one wants to test whether or not the
corrected background remains supersymmetric.  We found the simple
expression (\ref{Ddef}) for the corrected covariant derivative in the
gravitino transformation rule.  This expression, which is constructed
using the calibrating 4-form of the Spin(7) background, can be recast
in a purely Riemannian form, where no special tensors existing only in
special holonomy backgrounds are needed.  Remarkably, the Riemannian
expression, given in (\ref{riem1}), turns out to be identical to the
one first proposed in \cite{cfpss}, whose form was deduced from the
(considerably weaker) requirement of supersymmetry preservation for
corrected Calabi-Yau six-manifold compactifications.  Using the
corrected gravitino transformation rule, we illustrated with examples
the way in which one can derive corrected first-order equations for
metrics that have Spin(7) holonomy at leading order.

   We also extended our results to Spin(7) compactifications of M-theory,
This was considerably more complicated than the analysis at tree-level 
in string theory, partly because of the Chern-Simons terms that had 
to be taken 
into account and partly because of the topological constraint that forces
form fields to become non-vanishing when the Spin(7) manifold is compact (as 
implied by the term `compactification'). We gave a complete discussion 
of the 
corrections to (Minkowski)$_3\times K_8$ backgrounds, including for 
the first time a 
complete demonstration of supersymmetry preservation in the deformed
solutions. Our M-theory result implies a similar result  for one-loop 
corrected Spin(7) 
compactifications of IIA superstring theory.  It would be of interest 
to extend
this to the one-loop corrected IIB superstring theory, but we would 
not expect this
to introduce any essentially new features.

We also considered the case of (Minkowski)$_1\times K_{10}$
backgrounds in M-theory, where at leading order the manifold $K_{10}$
has a Ricci-flat K\"ahler metric with $SU(5)$ holonomy.  This case is
of particular interest because it probes features of M-theory that go beyond
those that can be directly accessed from perturbative string theory.
In order to avoid the complications arising from a topological constraint, 
we assumed that $H_8(K_{10})$ is trivial, which implies that $K_{10}$ is 
non-compact. Under this assumption, we were able to obtain equations 
for the corrections to the leading-order background. Remarkably, we found 
that the corrected $SU(5)$ holonomy backgrounds maintain their 
supersymmetry, 
assuming only that the previously-known correction term in the gravitino
transformation rule plays a r\^ole.  The corrected metric on $K_{10}$
is no longer K\"ahler, but it is still complex, with vanishing first Chern
class. Of course, it would be of considerable interest to extend 
these results to
compact  $K_{10}$. 

Finally, we wish to emphasise again the remarkable fact that the form
of the correction to the supersymmetry transformation rule first
proposed in \cite{cfpss} for string theory in the context of
six-dimensional Calabi-Yau compactifications continues to be
sufficient to guarantee supersymmetry preservation for
compactifications on Spin(7) manifolds. It is also sufficient  for Spin(7) 
compactifications, and certain $SU(5)$ `non-compactifications' of M-theory.
This suggests that it should be taken seriously as a candidate for the
complete gravitational part of the string or M-theory correction to
the gravitino supersymmetry transformation rule.

\section*{Acknowledgments}

   K.S.S. is grateful to the organisers of the {\it 37$^{\it th}$
International Ahrenshoop Symposium on the Theory of Elementary
Particles}, Berlin, August 23-27, 2004, for the invitation to talk on
the material presented in this paper, in a talk entitled {\it Quantum
Corrections to $G_2$ and Spin(7) Special Holonomy Spaces}.  C.N.P.,
K.S.S. and P.K.T. would like to thank the University of Barcelona, 
and C.N.P. and K.S.S. thank CERN, for
hospitality during the course of this work.  P.K.T. also thanks 
ICREA for financial support.

\end{document}